\documentclass[11pt]{article}

\usepackage[margin=1in]{geometry}

\usepackage[normalem]{ulem} 
\usepackage{siunitx}
\usepackage{upgreek}
\usepackage{graphicx}
\usepackage{amsmath}
\usepackage{amssymb}
\usepackage{xfrac}
\usepackage{soul}
\usepackage{pdfpages}
\usepackage{hyperref}
\usepackage{subcaption}
\usepackage{tabularx}
\usepackage{booktabs}
\usepackage{array}
\usepackage{bm}
\usepackage{mathptmx}

\usepackage[numbers,comma,sort&compress,square]{natbib}
\usepackage{titlesec}

\usepackage[font=footnotesize,labelfont=bf]{caption} 

\usepackage{float}
\usepackage[color=blue!15, size=footnotesize]{todonotes}
\usepackage{xcolor}

\titleformat{\section}{\Large\bfseries}{\thesection.}{1em}{}[]
\titleformat{\subsection}{\normalsize\bfseries}{\thesubsection.}{1em}{}[]
\titleformat{\subsubsection}{\normalsize\itshape}{\thesubsubsection.}{1em}{}[]
\titlespacing{\section}{0pt}{5pt}{0pt}[0pt]
\titlespacing{\subsection}{0pt}{3pt}{0pt}[0pt] 
\titlespacing{\subsubsection}{0pt}{0pt}{0pt}[0pt] 

\begin{document}

\begin{center}

\Large
\textbf{Shape-Independent Fluidity in Epithelial Cell Monolayers}\\

\vspace{11pt}

\normalsize
Pradip K. Bera$^{1}$, Anh Q. Nguyen$^{2,3}$, Molly McCord$^{1,4}$, Dapeng Bi$^{2,3}$, Jacob Notbohm$^{1,4,\dagger}$

$^1$ Department of Mechanical Engineering, University of Wisconsin--Madison, Madison, WI, USA\\
$^2$ Department of Physics, Northeastern University, Boston, MA, USA\\
$^3$ Center for Theoretical Biological Physics, Northeastern University, Boston, MA, USA\\
$^4$ Biophysics Program, University of Wisconsin--Madison, Madison, WI, USA
\vspace{11pt}

$^\dagger$ Correspondence: jknotbohm@wisc.edu

\end{center}

\vspace{11pt}

\section*{Abstract}
Tissue fluidity regulates biological processes such as embryonic development, wound healing, and cancer metastasis. In confluent epithelia, where cell packing fraction is effectively fixed, the prevailing paradigm postulates that fluidity is governed by a geometric shape index determined by the balance of cortical tension and intercellular adhesion. Here, we report that reducing cell-cell adhesion triggers an increase in fluidity with no change in cell shape index, cell density, substrate traction, or junctional line tension. The observed decoupling of shape and fluidity reveals that current vertex models, which treat adhesion as contributing solely to interfacial tension, are incomplete. To reconcile these findings, we extend the theoretical framework to account for the dual nature of adhesion---its thermodynamic role in setting interfacial adhesion energy at the cell-cell junctions and its kinetic role in generating viscous drag due to relative motion between adjacent cells. This generalized model quantitatively captures the experimental data, demonstrating that the interplay between adhesive energy and dissipative friction is essential for epithelial fluidity.

\section*{Introduction}

The ability of a tissue to deform and flow---referred to as tissue fluidity---underpins essential biological processes such as morphogenesis, wound healing, and cancer invasion~\cite{friedl2009collective, sadati2013collective, frittoli2023tissue, rinkevich2024tissue}. In epithelial monolayers, tissue fluidity is commonly framed as a collective mechanical property that emerges from cell-scale geometry and can control transitions between a jammed, solid-like state, wherein neighbor exchanges are suppressed, and an unjammed, fluid-like state, wherein cells can readily rearrange~\cite{herrera_kasza_2018biophysical_review, wang2020anisotropy, cai2022compressive}. Tissue fluidity promotes cell intercalation and epithelial remodeling during morphogenesis~\cite{machado2015emergent, tetley2018same}, facilitates epithelial wound healing in live \textit{Drosophila} wing imaginal discs~\cite{tetley2019tissue} and mouse skin epidermis~\cite{sarate2024dynamic}, and contributes to tissue regeneration and tubulogenesis~\cite{rinkevich2024tissue, ho2024role}. Given its central role in shaping tissue architecture and function, identifying the mechanisms that regulate fluidity is a fundamental question in tissue mechanics.

Current understanding of tissue fluidity focuses on a cell shape index $q$, defined as the ratio of perimeter to square root of area of each cell. When the average shape index is above a critical value, $q_c \approx 3.8$, tissues transition from solid-like to fluid-like behavior~\cite{bi2015density, sahu2020linear}. Within the fluid regime (\textit{i.e.}, for $q>q_c$), current understanding is that the shape index continues to be an indicator for fluidity, with higher shape index indicating greater fluidity. Physically, cell shape determines the geometric compatibility of cells within a confluent monolayer, where each cell tends to maintain its preferred, homeostatic area and perimeter. In this setting, the shape index quantifies how easily these constraints can be satisfied collectively. Rounded cells (with a lower shape index) are geometrically constrained. Any deformation typically requires a significant change in perimeter or area, which incurs an energetic penalty, leading to a rigid, solid-like response. In contrast, cells with a higher shape index are more elongated and possess greater geometric freedom: they can undergo shape fluctuations and small deformations without substantially changing their area or perimeter. This makes them mechanically more compliant, which enables the multicellular tissue to flow more readily. Experimental observations across diverse epithelial systems are broadly consistent with this picture: tissues composed of more elongated cells, characterized by a higher shape index, often exhibit enhanced motility, frequent neighbor exchanges, and collective flows~\cite{park2015unjamming, malinverno2017endocytic, mongera2018fluid, saraswathibhatla2020tractions, mitchel2020primary,wang2024cadherin, chisolm2025transitions}. 

These numerous experimental observations seem to position cell shape index $q$ as a unifying order parameter for epithelial fluidity. However, the predictive power of shape index relies on specific physical assumptions. In the quasistatic limit---wherein cells are jammed or nearly quiescent---shape is governed strictly by the balance of mechanical forces such as cortical tension, adhesion, and pressure~\cite{nagai2001dynamic, farhadifar2007influence, hufnagel2007mechanism, chiou2012mechanical}. Under these conditions, the tissue can be treated as a rate-independent system residing near an energy minimum, which is a fundamental premise of vertex-type models used to explain geometric-driven jamming. This framework fails to describe the system fully when cellular motility and dynamics become appreciable. In such regimes, cell shape reflects only the underlying rate-independent mechanical constraints; the shape index cannot fully capture rate-dependent effects such as dissipative forces due to relative motion between cells. Identifying the physical origins of such rate-dependent contributions, and understanding how they interact with the geometric controls captured by shape, is the central motivation of the present study. To this end, there is a need for approaches that go beyond the static energy landscape described by cell shape to explicitly account for the dynamics within the cell monolayer. Several extensions of the vertex-model framework have been developed to overcome this limitation, most notably the self-propelled Voronoi and active vertex models ~\cite{bi2016motility, pasupalak2020, sussman2018no, li2024relaxation, li2019mechanical}. In these approaches, tissue activity and dynamics are driven primarily by cell--substrate interactions, commonly modeled through substrate friction together with self-propulsion forces. In contrast, dissipative interactions between neighboring cells have been explored far less extensively. Although recent studies have introduced cell-cell or vertex--vertex viscous couplings into vertex-based models~\cite{arora2025viscous_important,rozman2025vertex}, the physical origin of these interactions and their influence on collective tissue dynamics remain poorly understood.

Here, we study collective migration within epithelial cell monolayers at constant number density focusing on collective cell motion within the fluid regime. We report substantial variation of fluidity that is independent of cell shape. This surprising result is caused by perturbing cell-cell adhesions, and we show that it is not caused by the line tension at the cell-cell interfaces nor the traction applied by the cells to the substrate. To explain how the adhesion-mediated fluidity can be independent of shape, we employ an extended vertex model that accounts for two effects of cell-cell adhesion, as follows. 
Firstly, adhesion is described physically by an interfacial surface energy, which describes the change in energy as cells adhere to one another. The interfacial surface energy is mathematically equivalent to adding a negative term to the effective junctional line tension, thereby acting against cortical contractility~\cite{brodland2002, lecuit2007cell, hilgenfeldt2008physical, paluch2009biology, manning2010coaction, winklbauer2015cell, lenne2021_tension_review}. Hence, adhesion reduces line tension, which softens cells and lowers the energetic barriers to deformation and neighbor exchange, facilitating rearrangements within a confluent layer~\cite{farhadifar2007influence, lecuit2007cell, cai2022compressive, perez2023tissue, park2015unjamming, saraswathibhatla2020tractions, mitchel2020primary, founounou2021tissue}.
Secondly, adhesion generates viscous drag at cell-cell interfaces, introducing a resistance that penalizes relative motion between adjacent cells~\cite{lenne2021_tension_review, fu2024regulation, arora2025viscous_important, mccord2025energy, mccord2026measurement}. 
Cell-cell adhesion, therefore, contributes to the physics in two ways, which we refer to, respectively, as an \textit{energetic} contribution and a \textit{dissipative} contribution. 
These two contributions have competing effects on tissue fluidity, which motivates close study of this competition. We develop an extended vertex model that accounts for both the energetic and dissipative contributions of adhesions. The extended model captures the experimental observations, namely the dissipative contribution of cell-cell adhesion enables shape-independent excursions in the rigidity phase diagram that are inaccessible in purely geometric descriptions.

\section*{Results}

\subsection*{Blocking E-cadherin-based cell-cell adhesion fluidizes cell monolayers without altering cell shape}

\begin{figure}[!thbp]
\centering
\includegraphics[width=6.5in]{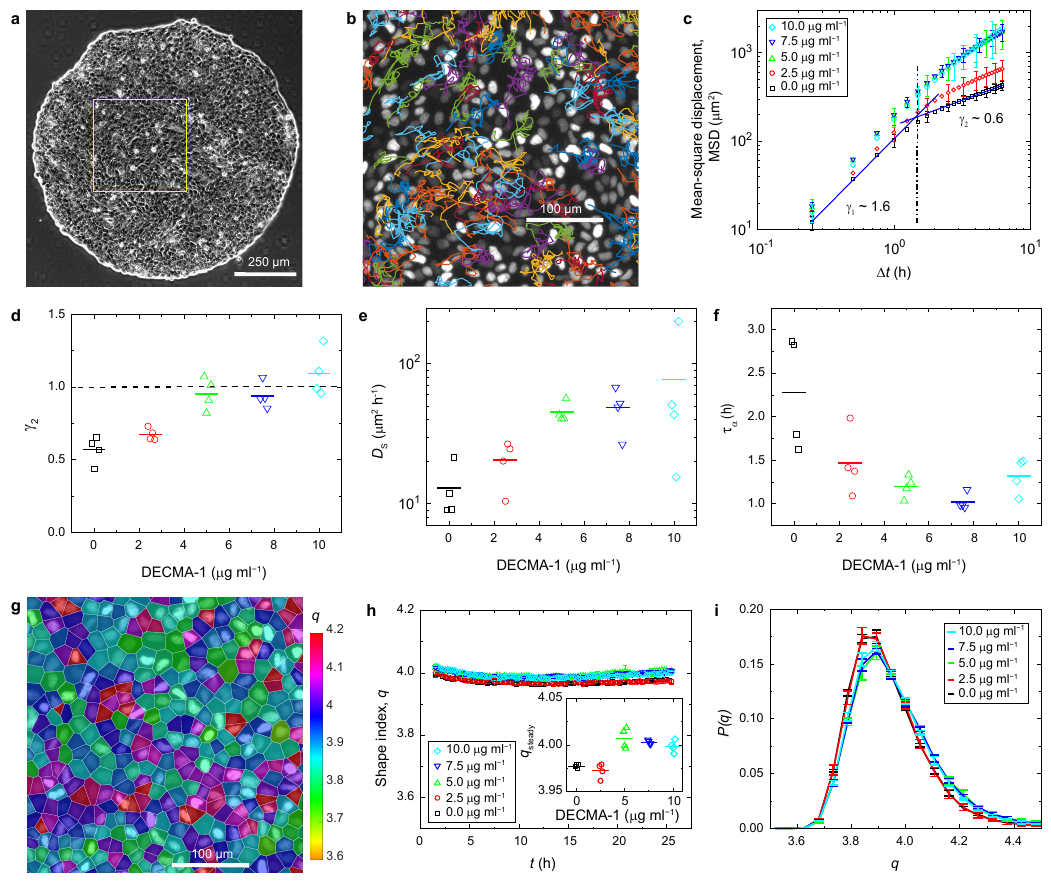}
\vspace{-6pt plus 2pt minus 2pt}
\caption{Fluidity of MDCK cell layers treated with the anti-E-cadherin antibody DECMA-1. (a) Phase-contrast image of a cell island, with the yellow boxed region magnified in later panels. (b) Representative trajectories overlaid on the nuclei image and color-coded to distinguish between different trajectories. (c) Mean-square displacement (MSD) calculated from steady-state trajectories ($t$ = 4~h to 24~h) for five DECMA-1 concentrations ($n$ = 4 per group; the same $n$ applies to panels d--f and h--i). Blue lines show fits to the 0~\textmu g~mL$^{-1}$ DECMA-1 condition over the $\Delta t$ (lag time) ranges 0.25--1.0~h and 2.5--6.25~h, yielding MSD exponents $\gamma_1$ and $\gamma_2$, respectively. Fits were extrapolated to estimate the superdiffusive-subdiffusive crossover time $\Delta t_c \approx$ 1.5~h, and length scale, $L_c \approx$ 17~\textmu m. (d--f) Scatter plots for individual islands at different DECMA-1 concentrations showing (d) long-time MSD exponent, $\gamma_2$, with dashed line indicating the subdiffusive boundary ($p$ = 8.5 $\times 10^{-7}$), (e) self diffusivity, $D_s$ in semi-log scale ($p$ = 0.014), (f) structural relaxation time, $\tau_\alpha$, extracted using DDM analysis ($p$ = 3.3 $\times 10^{-3}$). (g) Image of cell nuclei overlaid with Voronoi tessellations, color-coded by the cell shape index, $q$. (h) Average $q$ over time. Inset: steady-state shape index $q_\text{steady}$, defined as the average over $t$ = 23--24~h ($p$ = 1.1 $\times 10^{-3}$). (i) Probability distributions of steady-state shape index, $P(q)$. In panels c, h, and i, markers and error bars indicate means and standard deviations across independent cell monolayers. In all scatter plots, markers represent independent cell monolayers, and horizontal lines denote the group means. All $p$-values were obtained from two-sided linear correlation tests against the null hypothesis of no correlation.
}
\label{DECMA_Kinematic}
\end{figure}

Here, we aim to quantify tissue fluidity in confined monolayers of constant size. We seeded micropatterned islands of Madin-Darby canine kidney (MDCK) type II cells, approximately 1~mm in diameter, on polyacrylamide substrates with a Young's modulus of 6~kPa (Fig.~\ref{DECMA_Kinematic}a), following established protocols~\cite{notbohm2016cellular, bera2025traction}. To maintain a constant cell density, $\rho$, in the range of 2500 $< \rho <$ 3200~mm$^{-2}$ during our experiments, cell proliferation was inhibited using mitomycin C~\cite{jain2020role} after the desired density was reached (Supplementary Fig.~1). Cell-cell adhesion was disrupted by briefly exposing the monolayers to calcium- and magnesium-free phosphate-buffered saline (PBS). Subsequently, the reformation of adherens junctions was inhibited partially by adding the function-blocking antibody DECMA-1, which binds to E-cadherin and stabilizes it in a monomeric state, hence blocking E-cadherin-based cell-cell adhesion (Supplementary Fig.~2)~\cite{noren2001cadherin, shim2021overriding, chisolm2025transitions, mccord2026measurement}. We define the experiment time $t$ = 0~h as the time of DECMA-1 addition. Cell nuclei were tracked in two dimensions using time-lapse microscopy, enabled by the expression of green fluorescent protein fused to a nuclear localization signal (Fig.~\ref{DECMA_Kinematic}b and Supplementary Movies 1-5).

After reducing cell-cell adhesion, tissue fluidity was quantified using the mean-square displacement (MSD) calculated from steady-state cell trajectories acquired between $t$ = 4~h and 24~h (Fig.~\ref{DECMA_Kinematic}c). In untreated monolayers, the MSD exhibited a crossover from short-time superdiffusive dynamics, characterized by an exponent $\gamma_1$ = 1.6 (fitted over the lag time, $\Delta t$, in the range 0.25--1.0~h), to long-time subdiffusive dynamics with an exponent $\gamma_2$ = 0.6 (fitted range of $\Delta t$: 2.5--6.25~h). This crossover occurred at $\Delta t_c \approx$ 1.5~h, corresponding to a characteristic length scale $L_c \approx 17$~\textmu m. Notably, this length scale is comparable to the average cell diameter ($\approx$ 20~\textmu m), indicating that long-time dynamics are increasingly impeded by neighboring cells. 
With increasing DECMA-1 concentration, the long-time MSD exponent increases and then saturates for concentrations above 5.0~\textmu g~mL$^{-1}$ (Fig.~\ref{DECMA_Kinematic}d). 
We also computed the long-time self diffusivity $D_s$, defined as $D_s$ = $\lim_{t\to\infty} \text{MSD}(\Delta t)/(4\: \Delta t)$, which increases by more than a factor of six due to DECMA-1 (Fig.~\ref{DECMA_Kinematic}e). These trends indicate that blocking E-cadherin-based cell-cell adhesions enhances tissue fluidity.

To further characterize the dynamics, we applied differential dynamic microscopy (DDM)~\cite{ddm2025} to the phase-contrast image sequences to extract the structural relaxation time $\tau_\alpha$ (see Methods for details). 
Briefly, DDM is a coarse-grained measure and serves as a complementary, ensemble-level quantification of the underlying dynamics. Specifically, DDM extracts dynamics directly from image intensity fluctuations in Fourier space, without relying on individual particle tracking. 
Differential intensity patterns were computed at increasing time lags and azimuthally averaged, under the assumption of isotropic dynamics for each island. From these patterns, a time-dependent intermediate scattering function $F(k_0,\Delta t)$ was obtained using Eqn.~\ref{F_eqn} at a characteristic length scale $1/k_0$ = 17~{\textmu}m, set by the MSD crossover (Supplementary Fig.~3). For each island, the structural relaxation time $\tau_\alpha$ was defined as the $1/e$ decay time of the scattering function. Consistent with MSD-based measurements, $\tau_\alpha$ decreased with increasing DECMA-1 concentration (Fig.~\ref{DECMA_Kinematic}f), providing further confirmation that blocking E-cadherin binding substantially enhances tissue fluidity.

To explain these observations, we first turned to the shape index $q$, defined as the ratio of perimeter to square root of area for each cell. Prior models, supported by experiments, have shown that higher fluidity within the tissue coincides with a greater shape index~\cite{park2015unjamming, bi2016motility, mongera2018fluid, saraswathibhatla2020tractions,mitchel2020primary,  cai2022compressive, eckert2023, wang2024cadherin, chisolm2025transitions, kusaka2026cell}. 
To test whether shape index was elevated in response to reduced cell-cell adhesion, we quantified the average cell shape index $q$ using Voronoi tessellations (Fig.~\ref{DECMA_Kinematic}g). Across all experiments, cells exhibited relatively high aspect ratios, yielding values of $q$ well above the jamming transition threshold ($q_c \approx 3.8$), in agreement with previous studies~\cite{saraswathibhatla2020tractions}. Strikingly, despite the pronounced increase in tissue fluidity, reducing cell-cell adhesion did not produce a notable change in the cell shape index. Instead, the average shape index, $q$, barely increased for higher concentrations of DECMA-1 (Fig.~\ref{DECMA_Kinematic}h), and histograms of $q$ at steady state showed nearly identical distributions (Fig.~\ref{DECMA_Kinematic}i). To verify the Voronoi-based measurements, we also measured $q$ based on segmented cell boundaries obtained from phase contrast images of the cells, with results showing the same trends, namely $q$ was nearly unchanged in response to DECMA-1 (Supplementary Fig.~4).

To demonstrate that these results were not specific to the chosen cell density, we repeated these experiments in cell islands at a higher range of number density (3500 $< \rho <$ 4000~mm$^{-2}$). Results were consistent: higher DECMA-1 concentrations led to higher fluidity in cell motion without any change in the cell shape index $q$ (Supplementary Fig.~5).

\subsection*{DECMA-1 does not affect junctional line tension or traction forces}

The observed change in fluidity with no change in shape index goes against the current paradigm in the field. To identify the underlying mechanism, we investigated the effects of the DECMA-1 treatment on the cellular forces. First, we considered the junctional line tension at each cell-cell interface. It is commonly suggested that cell-cell adhesions act as a negative line tension~\cite{brodland2002, lecuit2007cell, farhadifar2007influence, paluch2009biology, manning2010coaction, winklbauer2015cell}, meaning that a reduction in adhesions by DECMA-1 would be expected to increase the junctional line tension. Alternatively, it is also possible that the effects of cell-cell adhesions are negligible compared to the cortical tension~\cite{krieg2008tensile, maitre2012adhesion, winklbauer2015cell}, in which case treatment with DECMA-1 would have no effect on the junctional line tension.

\begin{figure}[!thbp]
\vspace{-8pt plus 2pt minus 2pt}
\includegraphics[width=6.5in]{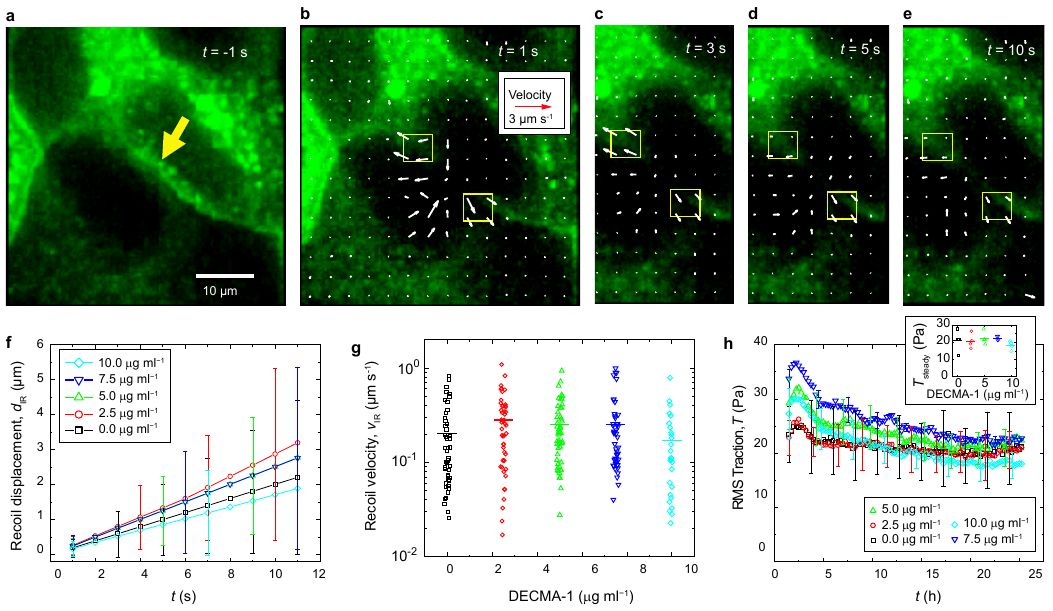}
\caption{Effects of DECMA-1 on cell-cell interfacial adhesion energy and cell-substrate traction. (a) Representative confocal image of MDCK monolayers with a live stain for F-actin; the yellow arrow indicates the junction segment 1~s, before ablation. (b-e) The same region after ablation at successive time points. Overlaid velocity vectors were obtained from Fast Iterative Digital Image Correlation (FIDIC), and yellow boxes denote regions of interest at opposite ends of the ablation site. The scale bar shown in (a) is the same for all panels (a)-(e). (f) End-to-end recoil displacement $d_\mathrm{IR}$ as a function of time, averaged over about 50 ablation sites for each DECMA-1 concentration. (g) Initial recoil velocity $v_\mathrm{IR}$, a proxy for junctional line tension, computed over the first 8~s after ablation; horizontal lines denote means ($p$ = 0.55, $n$ = 50, 49, 50, 50, 27 for each group, respectively; extended statistical test in Supplementary Table~1). (h) Root-mean-square (RMS) traction $T$ as a function of time; inset: steady-state traction $T_\text{steady}$, defined as the average over $t$ = 23--24~h ($p$ = 0.56, $n$ = 4 per group; extended statistical test in Supplementary Table~1). Error bars indicate standard deviations. Horizontal lines denote means of respective groups. All $p$-values were obtained from two-sided linear correlation tests against the null hypothesis of no correlation.}
\label{DECMA_Ablation}
\end{figure}

We probed junctional line tension using laser ablation experiments, wherein the junctions between neighboring cells were ablated, and the subsequent retraction rate was used as an indicator of the line tension. Snapshots from a representative experiment are shown in Fig.~\ref{DECMA_Ablation}a-e, wherein the cell edges on either side of an ablation recoiled over time. Under the assumption of constant viscosity, the short-time recoil velocity of the two ablated ends is proportional to the junctional line tension, whereas longer timescales reflect active remodeling processes. Hence, we quantified the recoil displacement over time (Fig.~\ref{DECMA_Ablation}f), which in turn allowed us to calculate the initial recoil velocity at each DECMA-1 concentration (Fig.~\ref{DECMA_Ablation}g). The initial recoil velocity $v_\mathrm{IR}$ spanned nearly two orders of magnitude (0.01--1~\textmu m~s$^{-1}$), reflecting substantial mechanical heterogeneity within the monolayer. The relationship between DECMA-1 concentration and recoil velocity was not monotonic: the recoil velocity in response to 2.5~\textmu g~mL$^{-1}$ DECMA-1 was slightly larger than the control, and the recoil velocities for all other concentrations were not statistically different from the control (Supplementary Table~1). There are two possible explanations for these observations. Firstly, E-cadherin may have had negligible effect on junctional line tension, which would be consistent with prior findings demonstrating important contributions of the cortex on line tension~\cite{krieg2008tensile, maitre2012adhesion, winklbauer2015cell}. Secondly, effects of E-cadherin could have been compensated by changes in cortical tension within each cell. No matter the underlying cause, the results show a negligible change in line tension, which does not explain the substantial effects of DECMA-1 on tissue fluidity.

Next, we considered that in response to the DECMA-1 treatment, the cells may have altered their force production, which can have a strong effect on fluidity in this system~\cite{saraswathibhatla2020tractions}. Therefore, we quantified the cell-substrate traction over time using traction force microscopy (Supplementary Fig.~6) ~\cite{butler2002traction, del2007spatio, trepat2009physical}. The root-mean-square (RMS) traction was nearly constant in time, and the steady state traction was not statistically different for different DECMA-1 concentrations (Fig.~\ref{DECMA_Ablation}h), ruling out traction as the reason for the DECMA-1-induced increase in fluidity. Hence, two of the major controllers of tissue fluidity and shape index identified in prior work, namely junctional line tension and traction~\cite{bi2015density, park2015unjamming, bi2016motility, saraswathibhatla2020tractions, mitchel2020primary}, do not explain the observations presented here.

\subsection*{Perturbing cell-cell adhesion by calcium chelation and E-cadherin knockout leads to shape-independent change in fluidity}

To independently perturb cell-cell adhesion through a mechanism different from antibody-mediated blocking, we performed complementary experiments using ethylene glycol tetraacetic acid (EGTA). Cadherin-based cell-cell adhesion requires extracellular calcium ions~\cite{shapiro2009structure}. 
EGTA chelates Ca$^{2+}$ in the culture medium, thereby weakening cadherin-mediated cell-cell adhesion. At EGTA concentrations $\geq$ 2~mM, monolayers lost confluency and cells detached from the substrate; therefore, all measurements were restricted to concentrations $\leq$ 1.8~mM. Here, the 0~mM EGTA condition served as a sham control, and $t$ = 0~h is the EGTA treatment time. We used a similar range of cell density as above (2600 $< \rho <$ 3300~mm$^{-2}$) (Supplementary Fig.~7).

\begin{figure}[!thbp]
\vspace{-4pt plus 2pt minus 2pt}
\includegraphics[width=6.5in]{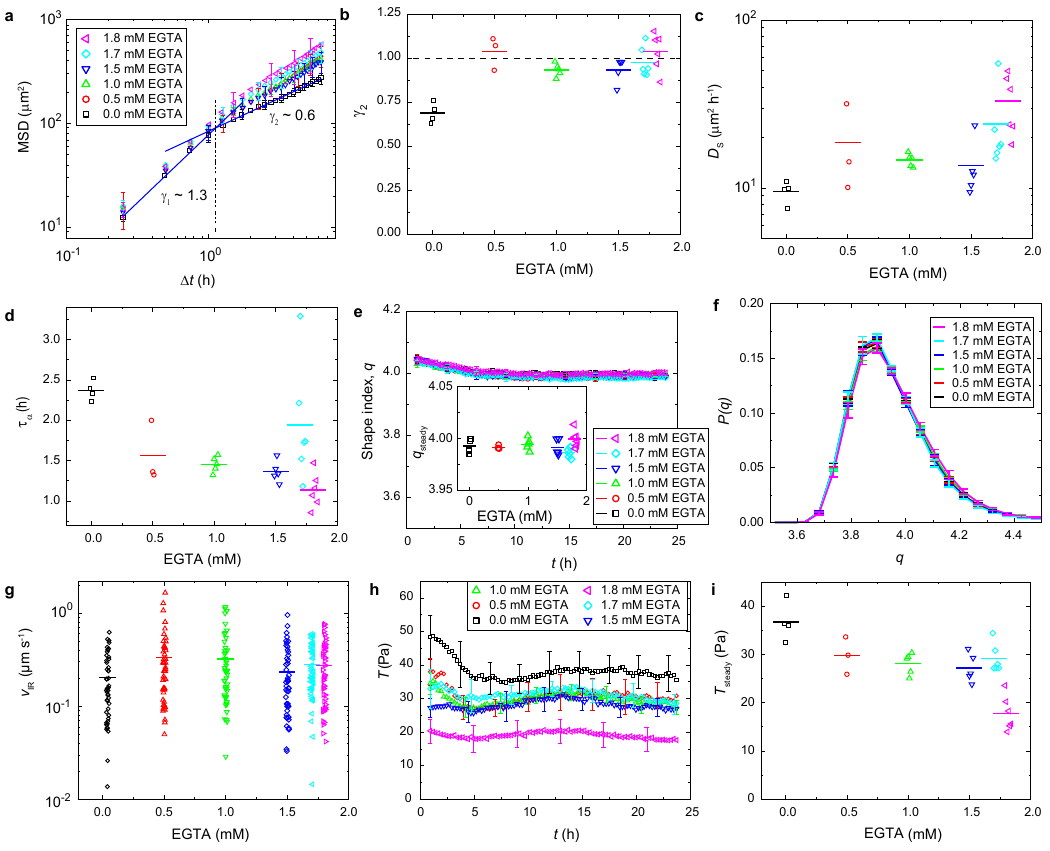}
\caption{Effects of EGTA on tissue fluidity, cell shape, junctional line tension, and traction. (a) MSD calculated from steady-state trajectories ($t$ = 7--24~h), for all EGTA conditions ($n$ = 4, 3, 5, 5, 6, 6 for each group, respectively; the same $n$ applies to panels b--f and h--i). Blue lines show fits to the 0~mM EGTA data in the $\Delta t$ intervals 0.25--1.0~h and 2--6.25~h, with MSD exponents $\gamma_1$ = 1.3 and $\gamma_2$ = 0.6, respectively, yielding $\Delta t_c \approx$ 1.1~h and $L_c \approx$ 10~\textmu m. (b--d) Scatter plots for each islands with EGTA showing (b) long-time MSD exponent, $\gamma_2$ ($p$ = 4.2 $\times 10^{-4}$), (c) self diffusivity, $D_s$ ($p$ = 8.2 $\times 10^{-3}$), and (d) structural relaxation time, $\tau_\alpha$ ($p$ = 0.012). (e) Average shape index, $q$, Inset: steady-state shape index $q_\text{steady}$, defined as the average over $t$ = 22--23~h ($p$ = 0.86; extended statistical test in Supplementary Table~2). (f) Steady-state distributions of $q$. (g) Initial recoil velocity, $v_\mathrm{IR}$, measured after laser ablation; horizontal lines indicate mean ($p$ = 0.63; $n$ = 50 per group; extended statistical test in Supplementary Table~2). (h) RMS traction, $T$, reaching steady state after $t \approx 7$~h. (i) Steady-state traction, $T_\text{steady}$ ($p$ = 1.1 $\times 10^{-4}$). In panels a, e, f, and h, markers and error bars indicate means and standard deviations across independent cell monolayers. In all scatter plots, markers represent independent cell monolayers, and horizontal lines denote the group means. All $p$-values were obtained from two-sided linear correlation tests against the null hypothesis of no correlation.} 
\label{EGTA_plots}
\end{figure}

Consistent with the DECMA-1 results, treatment with EGTA increases tissue fluidity, as evidenced by the increase in both the long-time MSD exponent $\gamma_2$ and the self diffusivity $D_s$ (Fig.~\ref{EGTA_plots}a--c). Furthermore, the structural relaxation time $\tau_\alpha$ decreased with increasing EGTA concentration, which also indicates that EGTA increases fluidity (Fig.~\ref{EGTA_plots}d).
Similar to the DECMA-1 treatment, distributions of shape index $q$ were unchanged by EGTA, which again indicates that the increase in fluidity was independent of cell shape (Fig.~\ref{EGTA_plots}e, f). 
Also similar to DECMA-1, laser ablation experiments show that the junctional line tension did not change monotonically with EGTA concentration (Fig.~\ref{EGTA_plots}g). Similar to the DECMA-1 experiments, the lower concentrations of EGTA (namely 0.5 and 1.0 mM) had retraction velocities that were statistically larger than control, but given the shape index $q$ remained constant, these changes in line tension apparently had negligible effects. 
The EGTA treatment decreased the cell-substrate tractions, especially at the highest concentration of EGTA, for which the RMS traction decreased by nearly a factor of two (Fig.~\ref{EGTA_plots}h, i). Interestingly, prior work using different treatments for cell force production showed that reducing traction causes a reduction in tissue fluidity~\cite{saraswathibhatla2020tractions}. The observation that fluidity was not reduced but rather increased in these experiments suggests that the effects of EGTA on the adhesions outweighed effects of traction.

Taken together, the EGTA experiments corroborate the DECMA-1 findings, demonstrating that reducing cell-cell adhesion enhances tissue fluidity without appreciably altering cell shape. Furthermore, even though EGTA also reduced traction, its effect on adhesions apparently outweighed effects of traction, indicating that cell-cell adhesions have an important contribution to fluidity. 
As a final test of the effects of reducing cell-cell adhesions on tissue fluidity, we performed co-culture experiments using MDCK wildtype (wt) cells mixed with E-cadherin knockout (KO) cells at ratios ranging from 0 (no KO cells) to 1 (all KO cells). As reported previously~\cite{balasubramaniam2021investigating}, the cell-substrate traction forces differed for KO compared to wt. Nevertheless, trends of the KO experiment supported results of the DECMA-1 and EGTA experiments, namely that reduced cell-cell adhesion led to an increase in tissue fluidity with negligible change in shape index (Supplementary Fig.~8).

\subsection*{An extended vertex model with dual roles of cell-cell adhesion captures shape-independent fluidity}

\begin{figure}[!htbp]
\centering
\includegraphics[width=1\linewidth]{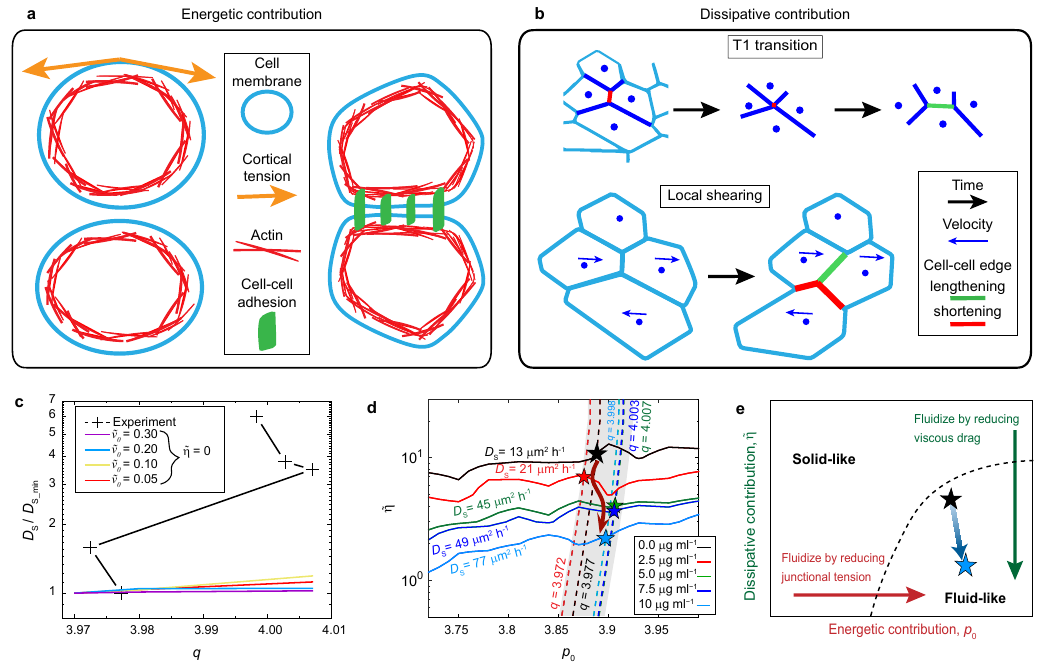}
\caption{Mapping shape-independent changes in fluidity in epithelial monolayers onto the vertex-model framework. (a) Schematic of the tension-related energetic contribution of cell-cell adhesion. Contractile tension generated by the actomyosin cortex tends to minimize cell perimeter and promote cell rounding, thereby disfavoring extended intercellular contacts. In contrast, cadherin-mediated adhesive interactions stabilize and extend cell-cell junctions, effectively reducing the junctional line tension and softening cells. 
(b) Schematic of the dissipative contribution of cell-cell adhesion. The top row shows a T1 transition, highlighting one cell-cell junction that shortens and vanishes and a new junction that forms and elongates. The bottom row depicts a form of local shearing, wherein some junctions lengthen and others shorten. Changes in junction length require formation and deformation of cell-cell adhesions and the cortical network, which leads to a viscous dissipation that is captured by $\tilde\eta$ in the extended vertex model. 
(c) Self diffusivity $D_s$ (normalized to its lowest measured value) as a function of shape index $q$, as observed in experiments and using the original vertex model for different values of $\tilde{v}_0$. The discrepancy between the curves from experiments and simulations demonstrates that the original vertex model does not capture the experimental observations. (d) Contours in the $(p_0,\tilde\eta)$ parameter space of the extended vertex model for self diffusivity $D_s$ (using mean values from Fig. \ref{DECMA_Kinematic}e) and cell shape index $q$ (using mean values from Fig. \ref{DECMA_Kinematic}h, inset). The background shading indicates experimental uncertainty in $q$ (defined as the standard deviation of $q$ across different cell monolayers). 
The intersections of the contours show the shape-independent trajectory through the ($p_0,\tilde\eta$) plane in response to DECMA-1-induced perturbations of cell-cell adhesion, as indicated by the brown arrow. (e) Conceptual phase diagram of tissue fluidity as a function of the energetic and dissipative contributions of cell-cell adhesion.}
\label{DECMA_modeling}
\end{figure}

To explain how the reduction of cell-cell adhesion increases tissue fluidity, we first describe how adhesions affect the tissue mechanics. As described above, there are two distinct yet interrelated mechanisms: an interfacial adhesion energy (energetic contribution) and a viscous drag (dissipative contribution).
The energetic contribution is mathematically equivalent to a line tension, which is caused by two competing effects in epithelial cells. As illustrated in Fig.~\ref{DECMA_modeling}a, cortical tension tends to minimize cell perimeter by rounding the cell~\cite{paluch2009biology, moazzeni2025n}. Conversely, the presence of cell-cell adhesion at intercellular contacts anchors neighboring cortices together, promoting the elongation of cell-cell junctions~\cite{maitre2011role, maitre2012adhesion}. Energetically, therefore, the effect of cell-cell adhesion can reduce the junctional line tension~\cite{brodland2002, lecuit2007cell, hilgenfeldt2008physical, paluch2009biology, manning2010coaction, winklbauer2015cell, lenne2021_tension_review}, 
making cells softer and lowering the energetic cost of deformation and topological rearrangement, which effectively promotes tissue fluidity. 
The dissipative contribution is illustrated in Fig.~\ref{DECMA_modeling}b, which shows a T1 transition and a local shearing deformation. For both cases, cell-cell junctions remodel by shortening or lengthening. The continual formation and deformation of these bonds and the cortical network to which they connect gives rise to an effective drag force localized at the cell-cell junctions, thereby dissipating mechanical energy. From a dynamical standpoint, adhesion impedes changes in length of the cell-cell junctions, imparting a viscous character to intercellular interactions~\cite{lenne2021_tension_review, fu2024regulation, arora2025viscous_important, mccord2025energy, mccord2026measurement}. 

While the energetic contribution of cell-cell adhesion is implemented and studied extensively in the vertex model framework~\cite{bi2015density, bi2016motility, sussman2018no, tetley2019tissue, sussman2020interplay, nguyen2025origin, ansell2025tunable}, the dissipative aspect has received much less attention~\cite{tong2023linear,rozman2025vertex,ray2024role}. Further, the competition between the two aspects of cell-cell adhesion has not been studied to date. In the vertex model, cells follow the overdamped equation of motion 0 = $\mathbf{F}_i^{\text{c-s}}+\mathbf{F}_i^\text{int}+\mathbf{F}_i^{\text{act}}$, where $\mathbf{F}_i^\text{c-s}$ = $-\mu \dot{\mathbf{r}}_i$ is the viscous force between cell $i$ and the substrate, with $\mu$ being the cell-substrate viscous coefficient, $\mathbf{r}_i$ being the position of cell $i$, and the over-dot denoting a time derivative. $\mathbf{F}^\text{int}_i$ = $-\nabla_{\mathbf{r}_i} E$ is the interaction force between cells arising from the tissue total energy function $E$ = $\sum_i \big [ K_A(A_i-A_0)^2+K_P(P_i-P_0)^2\big ]$, with $K_A$ and $K_P$ being the elastic moduli associated with area and perimeter deformations, respectively; $A_i$ and $P_i$ denote the instantaneous area and perimeter of cell $i$; and $A_0$ and $P_0$ are the corresponding preferred values, which are material properties set by cellular physiology. $\mathbf{F}_i^\text{act}$ = $\mu v_0\hat{n}_i$ is the active propulsion force produced by the cell with magnitude $v_0$ and polarity $\hat{\mathbf{n}}_i$ = $(\cos\theta_i,\sin\theta_i)$. 
The polarity angle $\theta$ undergoes rotational diffusion with zero mean and variance $2D_r$. The model can be made dimensionless using $\sqrt{\bar{A}}$ as the unit of length (with $\bar{A}$ being the average cell area), $\mu/(K_A \bar{A})$ as the unit of time, and $K_A\bar{A}^{3/2}$ as the unit of force. Note, after rescaling dimensional variables $x$ with these units, we use the nomenclature $\tilde{x}$ to indicate the corresponding dimensionless form. All model calculations are performed using dimensionless variables (see Methods for parameter estimation).

Using the original vertex model described above, we first sought to reproduce the experimentally observed relationship between the self diffusivity $D_s$ and the cell shape index $q$. We ran simulations over a range of values of the dimensionless preferred perimeter $p_0$ = $P_0/\sqrt{A_0}$ (with $A_0/\bar{A}=1$) and motility strength $\tilde{v}_0$ = $v_0\mu/(K_A\bar{A}^{3/2})$, which are the two primary control parameters governing tissue dynamics in the original vertex model, while keeping all other parameters fixed. For each simulation, the mean shape index was computed as $q$ = $\langle P_i/\sqrt{A_i}\rangle$ with the brackets indicating a mean over all cells. Additionally, the self diffusivity $D_s$ was measured in the simulations, allowing us to plot normalized $D_s$ as a function of $q$ for comparison to the experiments. Over the experimentally observed range of $q$, the experimental $D_s$ increased sixfold (Fig.~\ref{DECMA_modeling}c). In contrast, the original vertex model produces only a weak increase in $D_s$ across the experimentally observed $q$ range, with a maximum change of just 1.2-fold, which occurred for an intermediate value of $\tilde{v}_0$. This large quantitative mismatch demonstrates that the experimentally observed change in fluidity cannot be explained within the existing vertex model framework, in which changes in tissue dynamics are tightly constrained by changes in cell shape.

Motivated by the large changes in collective tissue fluidity that occur without appreciable changes in cell shape, we extend the original vertex model to explicitly incorporate both energetic and dissipative contributions of cell-cell adhesion, leading to a modified overdamped equation of motion, 
\begin{equation}
  0=\mathbf{F}_i^{\text{c-s}}+\mathbf{F}_i^\text{int}+\mathbf{F}_i^{\text{act}}+\mathbf{F}_i^{\text{c-c}}.
\end{equation}
Here, $\mathbf{F}^\text{c-c}$ is the drag force due to turnover in cell-cell adhesions and the cortex and has the form~\cite{Nguyen_CellCellAdhesion_DualRole_inprep}
\begin{equation}
  \mathbf{F}_{i}^{\text{c-c}}=-\sum_{j\in S_i}\eta(\mathbf{r}_{ij}\boldsymbol{\cdot} \mathbf{\dot{r}}_{ij})\hat{\mathbf{r}}_{ij},
\end{equation}
where $\eta$ is the junctional drag coefficient, $S_i$ is the set of neighboring vertices of vertex $i$, $\mathbf{r}_{ij}=\mathbf{r}_i-\mathbf{r}_j$ denotes the edge vector from vertex $j$ to vertex $i$, and $\hat{\mathbf{r}}_{ij}=\mathbf{r}_{ij} / |\mathbf{r}_{ij}|$ is the unit vector pointing from vertex $j$ to vertex $i$. The implementation of this yields the following dimensionless equation of motion:
\begin{equation}
\dot{\tilde{\mathbf{r}}}_i + \tilde{\eta} \sum_{j\in S_i}
\big[(\tilde{\mathbf{r}}_{ij}\boldsymbol{\cdot}\dot{\tilde{\mathbf{r}}}_{ij})\hat{\mathbf{r}}_{ij}\big]
= \mathbf{\tilde{F}}^{\text{int}}_i + \tilde{v}_0\hat{\mathbf{n}}_i,
\label{EoM}
\end{equation}
where $\tilde{\eta}$ = $\eta\sqrt{\bar{A}}/\mu$ is the dimensionless junctional drag coefficient, and $\mathbf{\tilde{F}}^{\text{int}}_i$ = $\mathbf{{F}}^{\text{int}}_i/(K_A\bar{A}^{3/2})$ is the dimensionless interaction force. In this extended vertex model, cell-cell adhesion controls two key parameters, $p_0$ and $\tilde{\eta}$, meaning that adhesion governs tissue fluidity by two competing mechanisms, namely the energetic and dissipative contributions described above. Interestingly, these two effects are in competition as follows. First, a reduction in adhesions would increase line tension (\textit{i.e.}, decrease $p_0$), which reduces fluidity. Secondly, reducing adhesion would reduce adhesive drag (\textit{i.e.,} reduce $\tilde\eta$), which enhances fluidity. As a result, tissue fluidity is no longer governed solely by geometric constraints but instead emerges from the interplay between energetic and dissipative contributions of cell-cell adhesion.

Representative animations from the simulations for different combinations of $p_0$ and $\tilde\eta$ are shown in Supplementary Movies 6-9. 
To enable quantitative comparison between simulations and experiments, we focus on two complementary observables: the self diffusivity $D_s$ and the dimensionless shape index $q$. The shape index $q$ provides a direct and scale-independent measure of tissue structure and can therefore be straightforwardly compared between experiments and simulations. The self diffusivity was normalized using the scales for length and time described above (see Methods for details on parameter estimation and Supplementary Table~3 for a summary of model parameters that best match the experiments).

Because our primary objective is to elucidate how cell-cell adhesion regulates tissue fluidity, we construct contour maps of both the dynamics (as quantified by the self diffusivity $D_s$) and shape index $q$, for which the two independent contributions of adhesion serve as control variables. Specifically, the dissipative (drag-related) component of cell-cell adhesion, quantified by the dimensionless parameter $\tilde\eta$, is used as one axis, while the energetic (tension-related) component, quantified by the preferred shape index $p_0$, is used as the other axis. These contour maps enable us to disentangle and assess the relative contributions of the energetic and dissipative aspects of cell-cell adhesion on tissue fluidity. We create contours that match $D_s$ and $q$ to the experimental data corresponding to each concentration of DECMA-1 in Fig.~\ref{DECMA_Kinematic}, thereby generating five contours of $D_s$ and five of $q$ in the $(p_0,\tilde\eta)$ plane (Fig.~\ref{DECMA_modeling}d). The intersections between contours of $D_s$ and $q$ identify the locations in the $(p_0,\tilde\eta)$ parameter space at which the model simultaneously matches both the tissue dynamics and the tissue geometry observed in each experiment. These intersections, therefore, provide our best estimate for mapping each experimental condition onto the contour map, enabling a comparison between experiment and theory. 
We note that $\tilde\eta$ is an inferred model parameter, which cannot be measured independently. As such, it is a coarse-grained representation of how dissipation within the cell monolayer changes in response to DECMA-1. 
The arrow in Fig.~\ref{DECMA_modeling}d indicates the trajectory through the parameter space along which tissue fluidity evolves as adhesion is systematically reduced by the DECMA-1 treatment. The results show that the dimensionless drag coefficient $\tilde\eta$ in the model decreases by nearly an order of magnitude, whereas the preferred shape index $p_0$ remains nearly constant, ranging from 3.87 to 3.91. We note that this change in $p_0$ is negligible compared to the variability in the shape index between different cell monolayers under the same DECMA-1 concentration (indicated by shading in Fig.~\ref{DECMA_modeling}d) and compared to the nearly order of magnitude change in $\tilde\eta$. Hence, in the model, changes in the dissipative contribution of adhesion are the major source of changes in fluidity. In the model, the interfacial adhesion energy part plays a negligible role, which is consistent with the minimal change in cortical tension observed in the experiments.

\section*{Discussion}

It has been widely accepted that cell shape, as quantified by the shape index, governs tissue fluidity. Our data show a striking change in fluidity that results from reducing cadherin-based cell-cell adhesions and is independent of shape index. Crucially, experiments were performed at constant density and constant cell-substrate traction. Laser ablation revealed no change in junctional line tension, indicating that the interfacial energy associated with cortical tension dominates over that of cell-cell adhesions. The original vertex model, which accounts for only the energetic effects of cell-cell adhesions and links tissue fluidity to the cell shape index, fails to explain our experimental observations. These findings indicate that cell-cell adhesions exert a second, purely dynamic effect on the collective motion: they generate forces that resist changes in length at cell-cell junctions, acting as an effective viscous dissipation~\cite{lenne2021_tension_review, fu2024regulation, arora2025viscous_important, mccord2025energy, mccord2026measurement}. We extended the original vertex model to incorporate both the energetic and dissipative contributions of cell-cell adhesion. Our extended vertex model, in which cell-cell adhesion controls two key parameters, the shape index and the viscous drag at cell-cell junctions, successfully reproduces the increase in tissue fluidity observed in the experiments.

An interesting observation in our experiments is that in control conditions, the dynamics were subdiffusive, as indicated by the long-time slope of MSD, $\gamma_2$, which could give the impression that control cell monolayers were in a jammed or solid-like state. However, the corresponding shape index was $q \approx 3.98$, which is above the threshold of $3.8$ for a solid-to-fluid transition identified in a prior vertex model~\cite{bi2015density}, implying that in control conditions the cells were in a fluid state. 
This finding is reminiscent of prior observations wherein partial epithelial-to-mesenchymal transition led to cells that were elongated and in the fluid state but moved slowly due to low motility~\cite{mitchel2020primary}. Similarly here, under control conditions, the cells were elongated ($q>3.8$), indicating a fluid, but the motion was sluggish, indicating a long viscoelastic relaxation time. Disrupting cell-cell adhesions led to faster motion, which is explained by a reduction in junctional viscosity. 
Our perturbations of cell-cell adhesions, therefore, do not drive the system across the solid--fluid boundary; instead, they shorten the relaxation time of a system that is already in the fluid regime, taking it from a slowly flowing state ($\gamma_2 \sim$ 0.6) to a more freely flowing one ($\gamma_2 \sim$ 1).

In many prior experimental studies of tissue jamming and fluidity, changes in cell-cell adhesion occur as part of broader perturbations that simultaneously alter additional variables, such as cell motility, cortical tension, cell-substrate adhesions, or number density. In such contexts, adhesion is either a downstream consequence of a larger regulatory program or a tuning parameter whose manipulation inevitably affects multiple mechanical and dynamical properties. By contrast, in our experiments, cell-cell adhesion is perturbed while other key factors influencing tissue fluidity remain effectively constant. This design isolates adhesion as the primary control parameter and directly implicates it as a driver of tissue fluidity. Interpreted through our extended vertex model, the results offer a unifying perspective on the relationship between adhesion and tissue unjamming. This framework helps reconcile the contradictory correlations between cell-cell adhesion and tissue rigidity previously reported in the literature~\cite{garcia2015physics, park2015unjamming, tetley2019tissue, cai2022compressive}. This perspective is summarized in a conceptual adhesion-driven jamming phase diagram (Fig.~\ref{DECMA_modeling}e). The horizontal axis represents the energetic contribution of adhesion: increasing adhesion can elevate the preferred shape index $p_0$ by reducing the effective junctional line tension, thereby promoting fluidity. The vertical axis captures the dissipative contribution: decreasing adhesion reduces intercellular viscous drag, which also enhances fluidity. Because these energetic and dissipative effects act in opposition, increasing cell-cell adhesion does not lead to a universally monotonic change in tissue fluidity; instead, the net outcome depends on which contribution dominates under a given condition. An additional implication of this framework is that changes in fluidity can occur without measurable changes in cell shape. A more detailed theoretical treatment of this dual and nontrivial role of adhesion is presented in a companion study \cite{Nguyen_CellCellAdhesion_DualRole_inprep}.

The shape-independent fluidity observed here upends the prevailing geometric paradigm that tissue fluidity is defined by cell shape, which presents new directions of study based on the effects of viscous dissipation. Future theoretical efforts will be needed to bridge the macroscopic dissipative framework presented here with the microscopic kinetics of intercellular bonds. For instance, incorporating shear-dependent bond rupture or non-linear junctional remodeling could reveal how local strain rates dynamically feed back into the effective junctional viscosity. Similarly, experiments will be needed that build on recent work~\cite{mccord2026measurement} to elucidate how viscous dissipation is related to the microstructure of the cells, including the density and location of cell-cell adhesions and the structure of the cytoskeleton. 
It will also be interesting to determine whether shape index and dissipation can be controlled independently, for example, by perturbing both cortical tension and cell-cell adhesion simultaneously, which could open up new opportunities for cancer treatment, tissue engineering, and wound healing.

\section*{Methods}

\subsection*{Cell culture}
MDCK type II cells expressing green fluorescent protein fused to a nuclear localization signal (a gift from Professor David Weitz, Harvard University) were cultured in Dulbecco's Modified Eagle's Medium (DMEM, 1~g~L$^{-1}$ glucose, 10-014-CV, Corning) supplemented with 10\% fetal bovine serum (Corning). 
MDCK E-cadherin knockout (KO) cells (from Professor Benoit Ladoux, Max-Planck-Zentrum für Physik und Medizin, Germany and described in ref. \cite{balasubramaniam2021investigating}) were cultured in DMEM (4.5~g~L$^{-1}$ glucose, 10-013-CV, Corning) supplemented with 10\% fetal bovine serum (Corning). 
Cells were maintained in an incubator (humidity maintained at 37$^\circ$C with 5\% CO$_2$) and were passaged regularly at approximately 70\% confluence. 

\subsection*{Substrate preparation and cell seeding}
All experiments were carried out with cells seeded as micropatterned islands of 1~mm diameter on 75~\textmu m-thick polyacrylamide (PA) gels with a Young’s modulus of 6~kPa, using the same cell culture medium. For the kinematics and traction experiments, 0.5~\textmu m-diameter fluorescent particles were centrifuged to the upper portion of the PA gel, as described in previous studies~\cite{notbohm2016cellular, bera2025traction}. 
A summary of substrate preparation and cell seeding is as follows. Glass-bottom dishes with 20~mm-diameter central wells and \#1.5 bottom cover glass were used as received from Cellvis. A freshly prepared PA solution (5.5\% w/v acrylamide and 0.2\% w/v bisacrylamide, Biorad Laboratories) was mixed with 0.03\% w/v fluorescent particles (0.5~\textmu m-diameter, carboxylate modified, Life Technologies) as required for traction force microscopy. Ammonium persulfate (0.08\% w/v, Biorad Laboratories) was added to initiate polymerization. Fifteen microliters of the gel (with or without beads) were deposited on the glass surface. Immediately after deposition, within a few minutes of polymerization, a coverslip was placed on top of the solution, and the gel was centrifuged upside down to localize the fluorescent particles near the upper portion of the gel. The final gel thickness was approximately 75~\textmu m. 

For micropatterning of 1~mm-diameter cell islands, polydimethysiloxane (PDMS, Sylgard 184) was cured into a 200~\textmu m-thick sheet, and holes of 1~mm diameter were punched using a biopsy punch to create a PDMS mask. The mask was sterilized with 70\% ethanol and incubated in 2\% w/v Pluronic F-127 (Sigma-Aldrich) overnight at room temperature. The treated PDMS mask was placed on each PA gel, which was then functionalized using sulfo-SANPAH (100~\textmu g~mL$^{-1}$, Pierce Biotechnology) to covalently crosslink rat-tail collagen I (Corning). One milliliter of 0.01~mg~mL$^{-1}$ collagen solution (in autoclaved DI water) was added per gel. Dishes were covered and kept overnight at 4$^\circ$C. 

200~{\textmu}l of cell suspension per dish were seeded at 1.6 $\times$ 10$^6$~mL$^{-1}$, and the cells were allowed to adhere for 2~h. The PDMS masks were then removed, the medium was refreshed, and cells were allowed to spread to confluence in each island over 8~h. For our co-culture experiments, nuclei of E-cadherin KO cells were stained by incubating the cells in the culture flasks with Hoechst 33342 dye (50~\textmu L~mL$^{-1}$, NucBlue, Thermo Fisher Scientific Inc.) for 1~h prior to seeding. Low-glucose medium (1~g~L$^{-1}$ DMEM with 10\% fetal bovine serum) was used to prepare two separate cell suspensions (wt and E-cadherin KO) having the same seeding concentration as above. The wt and KO cell suspensions were mixed according to the desired cell number ratio, $R$, and then 200{~\textmu}L of mixed suspension was seeded per dish.

\subsection*{Maintaining constant cell density}
In confined monolayers, cell density typically increases over time due to proliferation, which can obscure the effects of cell-cell adhesion, as tissue fluidity, cell shape, and cellular forces depend sensitively on density~\cite{kocgozlu2016epithelial, saraswathibhatla2020tractions, chisolm2025transitions}. Cell proliferation was inhibited using mitomycin C (10~\textmu g~mL$^{-1}$)~\cite{jain2020role}, thereby maintaining a constant cell density during experiments. After reaching the desired density, the culture medium was replaced with mitomycin C-containing medium, and cells were incubated for 1~h. After that, dishes were washed twice with fresh culture medium (without mitomycin C) and kept in the cell culture incubator with the fresh culture medium for about 12~h prior to imaging.

\subsection*{DECMA-1 treatment to decrease cell-cell adhesion}
To disrupt cell-cell adhesion, ten dishes containing PA gels with 1 mm cell islands were selected. The medium in each was replaced by PBS (Ca$^{2+}$, Mg$^{2+}$ free standard phosphate-buffered saline) for 30~min, thereby disrupting E-cadherin-mediated cell-cell adhesions~\cite{noren2001cadherin, shim2021overriding, chisolm2025transitions, mccord2026measurement}. Following this treatment, PBS was replaced with fresh culture medium supplemented with the desired concentration of the monoclonal E-cadherin antibody DECMA-1 (catalog no. 14-3249-82; research resource identifier AB-1210458; Thermo Fisher Scientific), thereby preventing the re-establishment of E-cadherin-mediated cell-cell adhesion. Note that the 0~\textmu g~mL$^{-1}$ DECMA-1 condition is sham control, in which all procedures were performed identically, except for the addition of DECMA-1. We define the experiment time $t$ = 0~h as the time of DECMA-1 addition.

\subsection*{Time-lapse imaging of cells and fluorescent particles}
Time-lapse imaging of cell islands was started within 1.5~h of the DECMA-1 treatments. Five of the ten dishes were used for time-lapse imaging. Phase contrast images, images of green fluorescent protein-labeled nuclei, Hoechst-labeled nuclei, and images of fluorescent particles embedded in the substrate were captured every 15~min using an automated microscope (Eclipse Ti, Nikon Instruments) with a $10\times$ numerical aperture 0.3 objective (Nikon) and an Orca Flash 4.0 camera (Hamamatsu) controlled by Elements Ar software (Nikon). Cells were maintained at 37$^\circ$C and 5\% CO$_2$ using a custom-built cage incubator on the microscope stage. After imaging, the cells were detached by replacing the cell medium with 0.05\% trypsin for 1~h, and images of the fluorescent particles in the substrate were captured for a traction-free reference state to compute cell-substrate tractions. The DECMA-1 time lapse imaging experiment was replicated independently three times at different cell number densities ($\rho$), yielding similar results.

\subsection*{Laser ablation experiments}
For laser ablation experiments, the other five dishes containing different concentrations of DECMA-1 were used. Live cells were stained for filamentous actin using the CellMask$^{TM}$ Orange Actin Tracking Stain (Thermo Fisher Scientific, A57244). After 12~h incubation, laser ablations were performed on a CSU-X spinning disk confocal microscope (Yokogawa) mounted on an Eclipse Ti base with a 60$\times$ NA 1.4 oil immersion objective (Nikon) and imaged with a Zyla camera (Andor), all controlled by the IQ3 acquisition software (Andor). Cell-cell edges were visualized and captured with the 561~nm laser line. A Micropoint laser (405~nm) was used to ablate cell-cell edges at 1.4~mW laser power (70\% of maximum laser power, with 200~\textmu J maximum pulses and 10~Hz repetition rate). A confocal image was captured 1~s before the ablation as a reference; immediately after the ablation, images were acquired at 1 frame per second for 10~s.

\subsection*{Fixed cell imaging for E-cadherin and DECMA-1 fluorescence}
The monoclonal antibody DECMA-1 binds to E-cadherin at cell-cell junctions. To verify this, cells were fixed and stained after 17~h of DECMA-1 treatment, with a cell density similar to that of the time-lapse experiments. Cells were washed twice with PBS and fixed for 20~min with 4\% w/v paraformaldehyde solution in PBS. Following fixation, cells were washed twice with PBS and permeabilized with 0.1\% v/v Triton X-100 in PBS. E-cadherin was stained overnight using a E-Cadherin (24E10) Rabbit mAB Alexa Fluor 488 Conjugate (Cell Signaling 3199S; 1:250 dilution), and DECMA-1 was stained using a Goat anti-Rat IgG (H+L) Cross-Adsorbed Secondary Antibody eFluor 647 (ThermoFisher A-21247; 1:250 dilution). A Nikon A1R confocal microscope (Nikon Instruments; 40$\times$-water immersion, NA 1.15) was used for imaging the samples and was controlled with NIS-Elements Ar software (Nikon). To visualize E-cadherin, and DECMA-1, a color-combined image was generated by merging color channels in Image-J to confirm the co-existence of E-cadherin and DECMA-1.

\subsection*{Data Analysis}
Fluorescent images of nuclei were segmented with the ImageJ plugin StarDist (version 0.3.0). Then, the `regionprops' function of MATLAB (version 2023b) was used to get centroid positions and trajectories of nuclei for each island. Centroids and trajectories were used for further analysis, including Voronoi diagrams, number density, MSD, and shape index. 

To enable measurements of shape index using actual cell boundaries, bright-field images of cell islands were segmented using Cellpose (version 3.0.7) \cite{stringer2021cellpose}. The area and perimeter of each cell were then calculated from the segmented boundaries using the MATLAB functions `bwboundaries' and `movmean' (window size 5~pix), with the latter applied for boundary smoothing to reduce curvature fluctuations.

For traction force microscopy, substrate displacements were computed by applying Fast Iterative Digital Image Correlation (FIDIC)~\cite{bar2015fast, franck_lab_2026_21514215} to images of the fluorescent particles using 32 $\times$ 32~pixel (21 $\times$ 21~\textmu m$^2$) subsets centered on a grid with spacing of 16~pixels (10.4~\textmu m). Traction was computed with unconstrained Fourier-transform traction microscopy accounting for substrates of finite thickness~\cite{butler2002traction, del2007spatio, trepat2009physical}. 

From the laser ablation data, the velocity field $v$ was calculated from confocal images of cells by using FIDIC using 32 $\times$ 32~pixel (3.5 $\times$ 3.5~\textmu m$^2$) subsets centered on a grid with spacing of 16~pixels (1.8~\textmu m) and dividing by the imaging time (1~s). To obtain end-to-end recoil velocity ($v_\mathrm{IR}$) or displacement ($d_\mathrm{IR}$) along the cell edge at the ablated region, two regions of interest (ROI) were manually identified on the opposite sides of the ablated hole.

\subsection*{EGTA treatment to reduce cell-cell adhesion} 
To reduce cell-cell adhesion, the calcium chelator EGTA (Sigma-Aldrich) was applied~\cite{saraswathibhatla2022coordination} instead of DECMA-1 while keeping all other aspects of the experimental workflow unchanged. A 100~mM stock solution of EGTA was prepared in PBS. After adding EGTA to PBS, the pH was slowly increased using 40\% (w/v) NaOH, as standard PBS (pH $\approx$ 7.4) cannot fully dissolve EGTA at 100~mM. After complete dissolution, the pH was readjusted to 7.4 using 40\% (v/v) HCl. The EGTA stock solution was sterile-filtered (0.22~\textmu m) and added to the cell culture medium at the final concentrations indicated. Here, the experiment time $t$ = 0~h is the EGTA treatment time. For 0~mM EGTA islands, an equivalent amount of PBS was added as a sham treatment. Time-lapse imaging of EGTA islands was started within an hour using the same settings as discussed above. 

\subsection*{Determination of the Structural Relaxation Time Using Differential Dynamic Microscopy (DDM)}

Differential Dynamics Microscopy (DDM)~\cite{ddm2025} is an imaging-based method that relies on time-lapse microscopy images to extract properties of system's dynamics. DDM extracts dynamical information by analyzing temporal fluctuations in real-space images, but quantifies this information in Fourier space through the structure function $D(\mathbf{k},\Delta t)$ = $\langle |\Delta I(\mathbf{k},t,\Delta t)|^2\rangle _t$, where $I(\mathbf{k},t,\Delta t)$ is the Fourier transform of the image difference $\Delta i(\mathbf{x},t,\Delta t)$ = $i(\mathbf{x},t+\Delta t)-i(\mathbf{x},t)$, with $i(\mathbf{x},t)$ being the image intensity (brightness). We applied DDM to our time-lapse phase contrast images of the cell islands to extract the normalized intermediate scattering function $F(k,\Delta t)$, which is related to azimuthal average $d(k,\Delta t)$ of the structure function~\cite{ddm2025}:
\begin{equation}
  F(k,\Delta t)=1-\frac{d(k,\Delta t)-B(k)}{A(k)}.
  \label{F_eqn}
\end{equation}
Here, $k$ = $|\mathbf{k}|$, $A(k)$ is the amplitude of the fluctuating signal, and $B(k)$ arises from the temporally uncorrelated detection noise. $B(k)$ is estimated as $d(k_\text{max},\Delta t\to\infty)$ with $k_\text{max}$ = $\pi/\delta_\text{px}$ being the upper wave vector limit that is set by the pixel size $\delta_\text{px}$ in our experiments. After getting $B(k)$, the signal fluctuation amplitude is estimated as $A(k)$ = $d(k,\Delta t\to \infty)-B(k)$. To extract the cellular rearrangement time, we evaluate the scattering function at the characteristic length scale set by the crossover in the MSD. At this length scale $1/k_0$ = 17~\textmu m, cell motion crosses over from independent, ballistic-like motion to diffusive-like motion governed by interactions with neighboring cells. The resultant $F_s(t)$ for different DECMA-1 concentrations show exponential decay (Supplementary Fig.~3). We define the structural relaxation time $\tau_\alpha$ as the time at which the scattering function decays to $1/e$. 
Consistent with the MSD results, $\tau_\alpha$ decreased with increasing DECMA-1 concentration (Fig. 1f). Beyond cross-validation, we report both metrics because they differ in experimental requirements and offer complementary advantages. MSD-based analysis requires reliable single-cell tracking, which in turn depends on fluorescent labeling (e.g., nuclear markers) and high-quality segmentation. In contrast, DDM operates directly on label-free images without segmentation or tracking, making it more readily applicable to systems where labeling is undesirable or tracking is unreliable (\textit{e.g.}, dense, three-dimensional, or unlabeled tissues).

\subsection*{Estimating model units and parameters}

At timescales shorter than the structural relaxation time $\tau_\alpha$, cell rearrangements are rare and cells are therefore weakly affected by interactions with their neighbors. In this regime, individual cell motion is primarily governed by single-cell motility rather than collective effects. Since the velocity persistence time also satisfies $1/D_r <$ 1~h $< \tau_\alpha$~\cite{saraswathibhatla2021}, the polarity of each cell is uncorrelated on a timescale that is shorter than the structural relaxation time. Consequently, on timescales $\Delta t < \tau_\alpha$, cell motion can be well approximated by that of a self-propelled particle undergoing rotational diffusion. We thus determine the parameters $v_0$ and $D_r$ by fitting the experimentally measured mean-squared displacement (MSD) at short times (specifically, for $\Delta t <$ 90~min) to the theoretical form: 
\begin{equation}
  \text{MSD}(\Delta t)=2\frac{v_0^2}{D_r^2}\big[\exp(-D_r\Delta t)+D_r\Delta t-1\big].
\end{equation}
The experimental MSD data were fitted to the theoretical expression (Supplementary Fig.~9). The theoretical model captures the dynamics at short timescales reasonably well, allowing us to estimate $v_0$ and $D_r$ as fitting parameters. The fitting results also suggest that the self-propulsion strength $v_0$ is minimally affected by DECMA-1 perturbation, \textit{i.e.}, $v_0 \approx$ 18~\textmu m~h$^{-1}$ for all DECMA-1 concentrations (Supplementary Fig.~9). Having obtained $v_0$, we next estimate the cell-substrate viscous coefficient $\mu$ using traction measurements together with the force balance relation:
\begin{equation}
  \mathbf{F}_i^{T}=\mathbf{F}_i^\text{c-s}+\mathbf{F}_i^\text{act}=\mu(v_0\hat n_i-\dot{\mathbf{r}}_i)
  \label{traction_force}
\end{equation}
Where $\mathbf{F}_i^T$ is the traction force exerted by the substrate on cell $i$. To compute $\mathbf{F}_i^{T}$ for each cell, we first extract the traction field defined on spatial grids over the cell island. We then perform a Voronoi tessellation using the cell nucleus positions to approximate individual cell domains. The traction acting on a given cell is obtained by vector summing the traction forces over all grid points that are mapped to the corresponding Voronoi region. Starting from Eqn.~\ref{traction_force}, we can eliminate the (unobserved) polarity vector $\hat{\mathbf{n}}_i$ by taking the norm of both sides, which yields $\|\mathbf{F}_i^T+\mu \dot{\mathbf{r}}_i\|$ = $\mu v_0$. This relation connects the measurable quantities $v_0$, $\mathbf{F}_i^{T}$ and $\dot{\mathbf{r}}_i$ to the unknown parameter $\mu$. We estimate the cell-substrate viscous coefficient $\mu$ by fitting this relation to the data. Specifically, we choose $\mu$ to minimize the residual $R(\mu)$ = $\sum_i \left( \big\lVert \mathbf{F}_i^{T} + \mu \dot{\mathbf{r}}_i \big\rVert - \mu v_0 \right)^2$. We determine $\mu$ from the 0~\textmu g~mL$^{-1}$ DECMA-1 condition and use this value for all other DECMA-1 concentrations, assuming that DECMA-1 does not significantly affect cell--substrate interactions, as indicated by the weak dependence of $v_0$ on DECMA-1 concentration. All simulations, for both the original and extended vertex model, are performed in a square domain with periodic boundary conditions and with $A_0/\bar{A}=1$. The parameter $K_A$ is obtained from the areal strain modulus $\kappa_A$ = 0.05~N~m$^{-1}$ for MDCK II cells, which was estimated using Atomic Force Microscopy (AFM)~\cite{bodenschatz2022epithelial}, as $K_A$ = $\kappa_A / \bar{A} \approx$ 0.11~N~m$^{-3}$. Using these estimated parameters, we express our experimental results in units of length $l_0$ = $\sqrt{\bar{A}} \approx$ 20~\textmu m and time $t_0$ = $\mu / (K_A \bar{A}) \approx$ 0.0055~h.

In these length and time units, a value of $v_0 \approx$ 18~\textmu m~h$^{-1}$ corresponds to a dimensionless $\tilde{v}_0$ = 0.0027, which is practically 0 for our simulation, making the dynamics difficult to be describe in a reasonable simulation time. Exploiting the proportionality between the self diffusivity $D_s$ and the persistence length $L_p$ = $v_0/D_r \approx$ 0.3$l_0$, we run simulations at larger $\tilde{v}_0$ and $\tilde{D}_r$ while keeping their ratio fixed to match the experimental persistence length. Specifically, we choose $\tilde{v}_0$ = 0.3 and $\tilde{D}_r$ = 1, which preserves the single-cell persistence length $\tilde{L}_p$ = 0.3 while accelerating convergence to the long-time diffusive regime. Importantly, this choice of motility parameters does not affect our central conclusion regarding the dominant role of dissipative adhesion. While increasing $\tilde{v}_0$ enhances fluidity, its effect can be offset by a corresponding increase in the viscous drag cell-cell adhesion $\tilde{\eta}$, which is equivalent to a vertical shift in the $p_0-\tilde{\eta}$ plane. Such a shift leaves the direction and qualitative trend of the fluidity trajectory unchanged. We therefore expect the dominance of dissipative adhesion to be robust to the specific choice of $\tilde{v}_0$ and $\tilde{D}_r$, provided $L_p$ is held fixed. To generate the contours of constant $D_s$ and constant $q$, we systematically sweep the parameter space by varying $p_0$ from 3.72 to 3.99 and $\tilde{\eta}$ from 0.01 to 100, while keeping $\tilde{K}_P$ = $K_P/(K_A\bar{A)}$ fixed at a value of 1. For each parameter pair $(p_0,\tilde{\eta})$, we simulate the tissue dynamics with 400 polygons by numerically integrating Eqn.~\ref{EoM} using an explicit Euler scheme with time step $dt$ = 0.05, and we evolve the system until a total simulation time of $10^4$ is reached, ensuring that steady-state behavior is attained. 

Accurately reproducing the full experimental geometry, system size, and boundary conditions within the vertex model framework is computationally expensive. Additionally, such detailed implementations often introduce significant complexity without providing proportional insight into the underlying mechanisms. Our goal was therefore not to exactly replicate the experimental configuration, but to estimate model parameters based on experimentally accessible measurements. We used periodic boundary conditions in both dimensions as a simplifying modeling choice, which we justify as follows. We estimate the magnitude of potential boundary effects in the experiments as follows. The largest steady-state self diffusivity observed in our experiments is $D_s \approx$ 77~\textmu m$^2$~h$^{-1}$ (Fig. 1e), corresponding to a typical cell displacement over the 24~h experimental window of ${(4 D_s t)}^{1/2} \approx (4\times77\times24)^{1/2}$~\textmu m $\approx$ 86~\textmu m ($\approx 4$ cell diameters). This represents an upper bound on how far cells near the island boundary could influence the bulk dynamics. For our 1~mm-diameter islands (radius 500~\textmu m), the annular region within this distance of the boundary contains approximately $1 - (414/500)^2 \approx 30$\% of the cells, with the remaining 70\% residing in a bulk-like environment. For all other (less fluid) conditions, this fraction is substantially smaller. Because the observables we focus on---MSD, self diffusivity, structural relaxation time, and shape index---are bulk-averaged quantities, we expect them to be insensitive to boundary effects. Our simulations consist of $\sim$ 400 polygons ($\sim$ 20 polygons along each direction) with periodic boundary conditions. The simulation box dimensions ($\sim$ 20 polygons × 20 polygons) are significantly larger than the characteristic cell displacement ($\sim$ 4 polygons). Therefore, the system size is sufficient to capture the experimentally observed trends reliably.

\subsection*{Statistics and reproducibility}
The experiments tested the response to a range of concentrations of DECMA-1 or EGTA and a range of mixing ratios between wt and E-cadherin KO cells. For these experiments, statistical tests using categorical independent variables, such as ANOVA, do not accurately capture the effect of concentration or mixing ratio, which are continuous variables. Therefore, for each data set, we began with a two-sided linear correlation, and acquired the $p$ value against the null hypothesis of no correlation, with statistical significance defined as $p < 0.05$. 
For all cases in which the correlation returned $p > 0.05$, we performed a more detailed investigation, using one-way ANOVA with Dunnett's test, which compared the data from each concentration of DECMA-1 or EGTA to the sham control. $p$ values from the Dunnett's test are included in the supplementary information.

The time-lapse imaging and laser ablation experiments involving DECMA-1, EGTA treatments, and wt-KO co-culture were repeated multiple times to assess reproducibility. All groups were subjected to identical culture, treatment, and imaging protocols to control for potential covariates, thereby ensuring comparability across conditions. The results shown in the manuscript are from one of each of the multiple different experiments. Importantly, the observed trends were consistent across all repeated experiments.

\section*{Data availability}
The source data for Fig.~\ref{DECMA_Kinematic}c-f,h-i, Fig.~\ref{DECMA_Ablation}f-h, Fig.~\ref{EGTA_plots}a-i, Fig.~\ref{DECMA_modeling}c-d, and for Supplementary Fig.~1, Fig.~3, Fig.~4b, Fig.~5, Fig.~7, Fig.~8, Fig.~9 are provided as a Source Data file with this paper. The raw data supporting the findings of this study are freely available in the Dryad Digital Repository (\cite{bera2026shape2dryad}) under the Creative Commons CC0 Public Domain Dedication. Data from replication experiments are available from the corresponding author upon reasonable request.

\section*{Code availability}
Custom code used to quantify the displacement fields from experimental images, FIDIC~\cite{franck_lab_2026_21514215}, and to calculate cell-substrate traction, Cell-Traction-Stress~\cite{jacob_notbohm_2026_21514235}, are publicly available. The custom code developed for the simulation in this study is available in Zenodo repository~\cite{nqanh1995_2026_21536426}.




\section*{Acknowledgments}
We thank Christian Franck for the use of the A1R confocal microscope. 

\section*{Funding Statement}
J.N. acknowledges support from NSF Grant No. CMMI-2205141 and NIH Grant No. R35GM151171. A.N. and D.B. acknowledge support from NSF Grants No. DMR-2046683 and PHY-2019745, the Sloan Research Fellowship, NIH Grant No. R35GM150494 and the Human Frontier Science Program research grant (RGP0007/2022). 

\section*{Author Contributions Statement}
P.K.B. performed the experiments and data analysis under the supervision of J.N. A.Q.N. performed the simulations under the supervision of D.B. M.M. performed fixed-cell confocal imaging. P.K.B., A.Q.N., D.B., and J.N. conceived the study, interpreted the results, and wrote the manuscript.

\section*{Competing Interests Statement}
The authors declare no competing interests.

\section*{Additional information}
Supplementary information is available for this paper and includes Supplementary Movies, Supplementary Figures, and Supplementary Tables.
Correspondence and requests for materials should be addressed to Pradip K. Bera or Jacob Notbohm.

\newpage

\begin{center}
\large\centering \textbf{Supplementary Information}

\Large
\textbf{Shape-Independent Fluidity in Epithelial Cell Monolayers}\\

\vspace{11pt}

\normalsize
Pradip K. Bera$^{1}$, Anh Q. Nguyen$^{2,3}$, Molly McCord$^{1,4}$, Dapeng Bi$^{2,3}$, Jacob Notbohm$^{1,4,\dagger}$

$^1$ Department of Mechanical Engineering, University of Wisconsin--Madison, Madison, WI, USA\\
$^2$ Department of Physics, Northeastern University, Boston, MA, USA\\
$^3$ Center for Theoretical Biological Physics, Northeastern University, Boston, MA, USA\\
$^4$ Biophysics Program, University of Wisconsin--Madison, Madison, WI, USA
\vspace{11pt}

$^\dagger$ Correspondence: jknotbohm@wisc.edu

\end{center}

\vspace{.4in}
\noindent Contents:\\
\indent Supplementary Movie Details\\
\indent Supplementary Figures 1-9\\
\indent Supplementary Tables 1-5

\newpage
\section*{Supplementary Movie Details}

Supplementary Movies 1--5 are time-lapse images of MDCK cell monolayers in 1 mm islands. Fluorescent images of cell nuclei (green channel) and phase-contrast images (gray channel) were color-merged for improved visualization. Each movie corresponds to a different DECMA-1 concentration.

Supplementary Movies 6--9 are representative animations of the extended vertex model for different combinations of preferred shape index ($p_0$) and dimensionless drag coefficient ($\tilde\eta$), keeping other parameters fixed. To visualize the dynamics, the color of each vertex is maintained throughout all frames.

\clearpage

\section*{Supplementary Figures}

\renewcommand{\figurename}{Supplementary Figure}
\setcounter{figure}{0}

\begin{figure}[!htbp]
\centering
\includegraphics[width=6.5in]{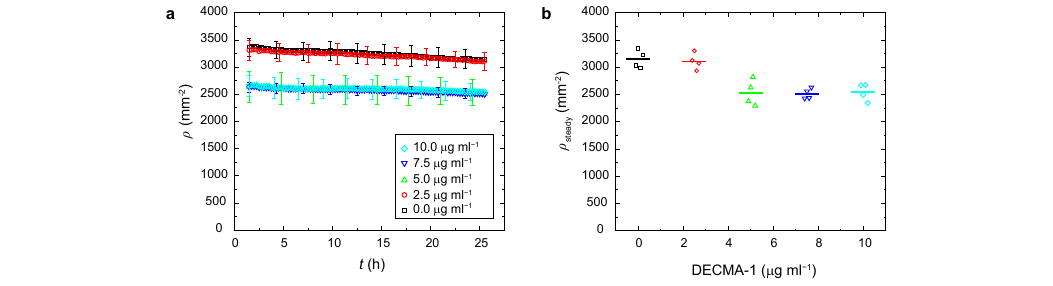}
\caption{Number density during the DECMA-1 experiment (Fig.~1). Number density was computed at each frame by counting nuclei within each island and dividing by island area. (a) Average number density $\rho$ for each DECMA-1 concentration, with error bars indicating standard deviations across different cell monolayers ($n$ = 4 per group; the same $n$ applies to panel b). (b) Plotted $\rho_\text{steady}$, the steady state density defined as the average over $t$ = 23--24~h ($p$ = 5.8 $\times 10^{-5}$, two-sided linear correlation test against the null hypothesis of no correlation). Horizontal lines denote means of respective groups. Although the number density $\rho$ was statistically different across the different DECMA-1 concentrations, the relative values are in a small range of $\approx$ 2500--3200 mm$^{-2}$. Prior studies that have shown effects of density on collective migration~\cite{saraswathibhatla2020tractions} have used a much larger range, \textit{e.g.}, 1200--4200 mm$^{-2}$, meaning that any effects of the small range of density in these experiments are likely to be modest. Furthermore, additional experimental data in this manuscript (main text Fig.~3 and Supplementary Fig.~5) show no correlation between density and kinematics, which rules out density as the cause of the change in fluidity reported here.
}
\label{Num_density}
\end{figure}

\begin{figure}[!htbp]
\centering
\includegraphics[width=6.5in]{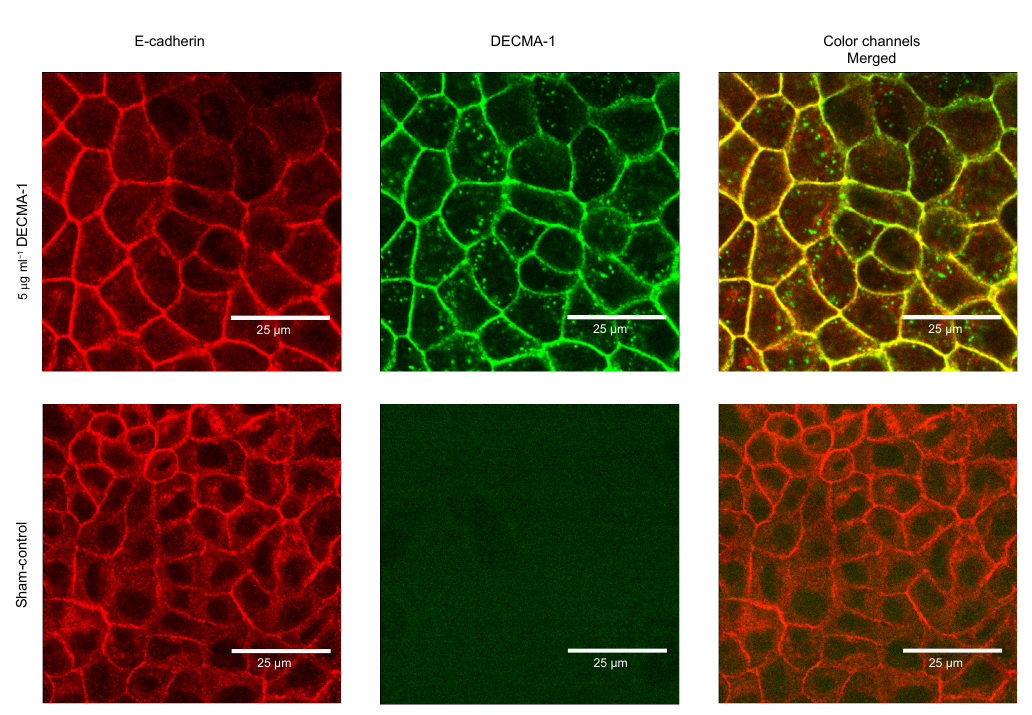}
\caption{Verification that DECMA-1 binds to E-cadherin. Top row, left to right: Confocal images of E-cadherin (red) and DECMA-1 (green) in cells fixed after 17~h, for a represented region in a cell monolayer treated with 5~\textmu g~ml$^{-1}$ DECMA-1; merging color channels appears yellow, indicating co-localization of E-cadherin and DECMA-1, which verifies that DECMA-1 bound to E-cadherin. Bottom row, left to right: similar plots for 0~\textmu g~ml$^{-1}$ show no green fluorescence, indicating that the antibody used for imaging DECMA-1 bound specifically to DECMA-1 in the 5~\textmu g~ml$^{-1}$ condition. }
\label{confoc_imgs}
\end{figure}

\begin{figure}[!htbp]
\centering
\includegraphics[width=6.5in]{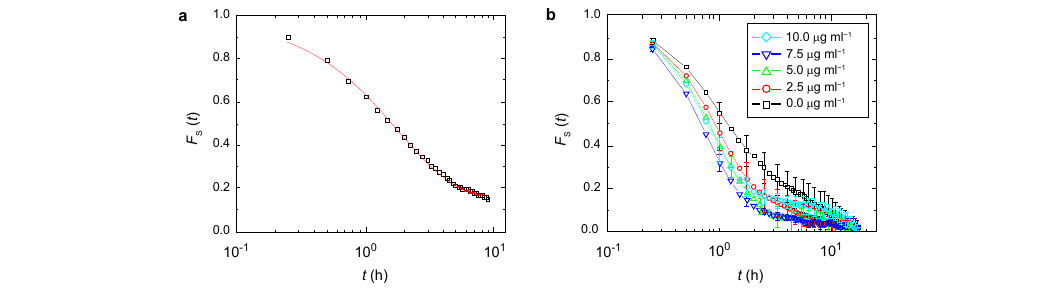}
\caption{Self-intermediate scattering function for different concentrations of DECMA-1. (a) For a representative 0~\textmu g~ml$^{-1}$ DECMA-1 island, the self-intermediate scattering function $F_s$ is plotted in time; the red curve is an exponential fit. (b) $F_s$ is plotted for all the DECMA-1 concentrations. $F_s$ decays more rapidly at higher DECMA-1 concentrations, indicating smaller $\tau_\alpha$. Markers and error bars indicate, respectively, averages and standard deviations across different independent cell islands ($n$ = 4 per group).}
\label{Fs_qt}
\end{figure}

\begin{figure}[!htbp]
\centering
\includegraphics[width=6.5in]{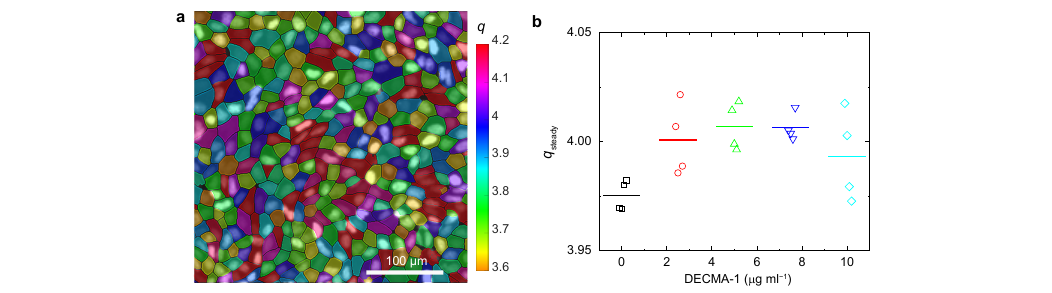}
\caption{Measurements of the shape index using segmented cell boundaries are consistent with the Voronoi tessellation analysis and confirm the negligible variation in shape index in response to DECMA-1. Phase-contrast images were segmented using the Cellpose software (main text Methods). The shape index, $q$, was calculated for each cell from the segmented outlines. (a) Representative image of Cellpose-segmented cell outlines, colored by the corresponding values of $q$, overlaid on the corresponding image of cell nuclei. The region shown here is the same as that in Fig. 1g of the main text. (b) Steady-state shape index, $q_\text{steady}$, defined as the average over $t$ = 15--17~h for each island ($p$ = 0.13, two-sided linear correlation test against the null hypothesis of no correlation with $n$ = 4 per group; extended statistical test in Supplementary Table~4). Horizontal lines indicate the means.}
\label{Cellposed_DECMA}
\end{figure}

\begin{figure}[!htbp]
\centering
\includegraphics[width=6.5in]{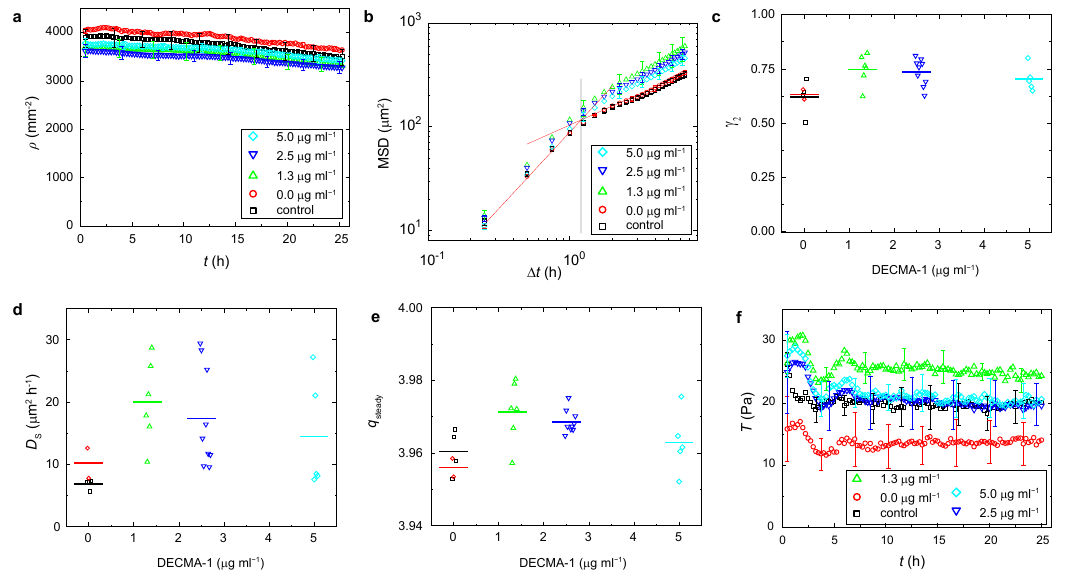}
\caption{Effects of DECMA-1 on kinematics and forces in cell islands at higher cell density (3500 $< \rho <$ 4500~mm$^{-2}$). Control monolayers had no PBS added, and the 0~\textmu g~ml$^{-1}$ concentration of DECMA-1 is a sham control, wherein PBS was added followed by replacing the PBS with cell culture medium containing no DECMA-1. Plots for: (a) number density $\rho$ ($n$ = 4, 2, 6, 9, 5 for each group, respectively; the same $n$ applies to panels b–f), (b) mean-squared displacement (MSD), (c) the long-time MSD exponent $\gamma_2$ ($p$ = 0.040), (d) self diffusivity $D_s$ ($p$ = 0.35), (e) steady-state cell shape index $q_\text{steady}$ ($p$ = 0.68) considering time interval $t$ = 23--24~h, and (f) traction $T$. Error bars are standard deviations. Horizontal lines denote means of respective groups. Extended statistical tests are in Supplementary Table~5. All $p$-values were obtained from two-sided linear correlation tests against the null hypothesis of no correlation.}
\label{DECMA_Repeated}
\end{figure}

\begin{figure}[!htbp]
\centering
\includegraphics[width=6.0in]{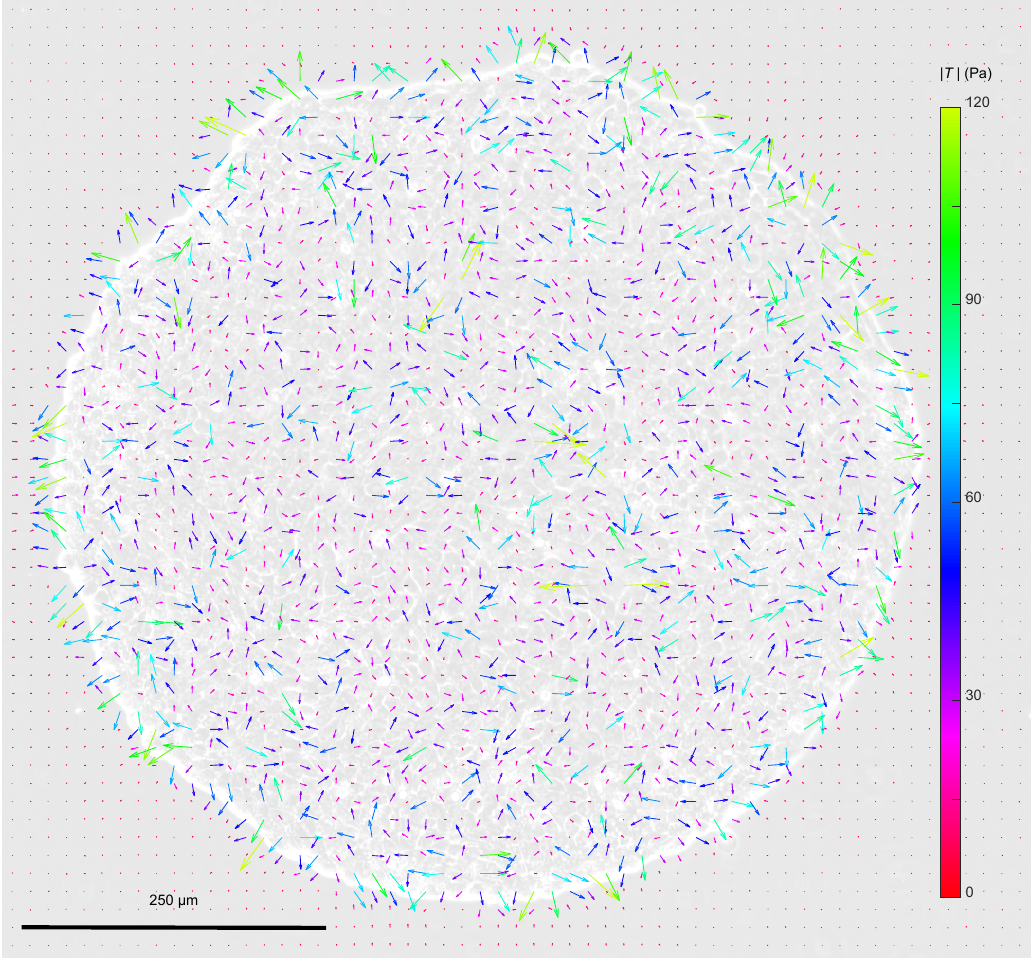}
\caption{Representative two-dimensional traction vector field (corresponding to Fig.~1a in the main text). Vectors are spatially downsampled by a factor of two for clarity, color-coded by traction magnitude, and overlaid on a semi-transparent monolayer image to enhance visualization. Small arrows at grid points far from the island indicate the traction noise level.}
\label{TFM}
\end{figure}

\begin{figure}[!htbp]
\centering
\includegraphics[width=6.5in]{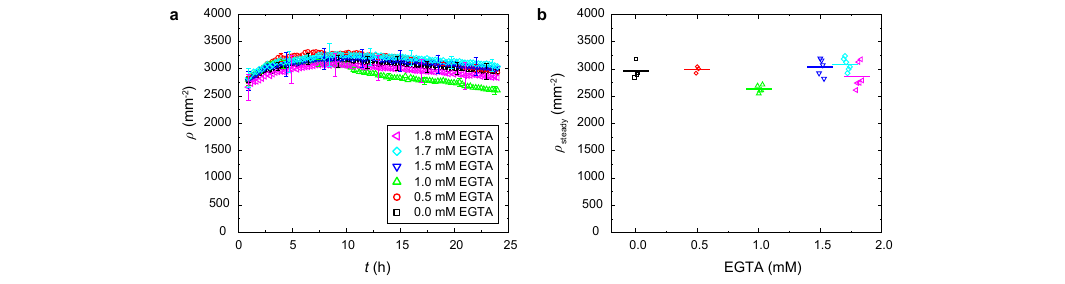}
\caption{Number density in EGTA experiment. (a) Number density $\rho$ plotted in time ($n$ = 4, 3, 5, 5, 6, 6 for each group, respectively; the same $n$ applies to panel b). (b) $\rho_\text{steady}$ steady state density, defined as the average over $t$ =22--23~h ($p$ = 0.58, two-sided linear correlation test against the null hypothesis of no correlation; extended statistical test in Supplementary Table~2).}
\label{EGTA_SI_plots}
\end{figure}

\begin{figure}[!htbp]
\centering
\includegraphics[width=6.5in]{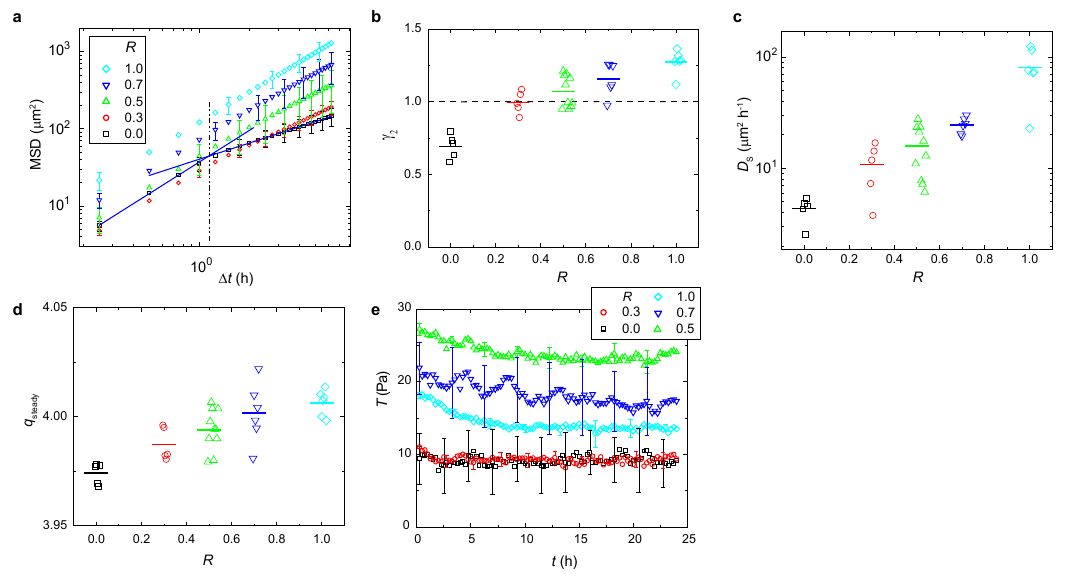}
\caption{Results of co-culture experiments using mixed populations of MDCK wildtype (wt) and E-cadherin knockout (KO) cells. Experiments were performed at varying number ratio of KO cells, defined as $R$ = $\rho_{KO}/\rho$ with $\rho_{KO}$ being the number density of KO cells and $\rho$ being the number density of all cells, which was in the range $4000 < \rho < 5700$~mm$^{-2}$. Experiments were performed with cell islands each having one of five different ratios $R$ ranging from 0 (meaning no KO cells) to 1 (meaning all KO cells). (a) MSD vs $\Delta t$ for different values of $R$ ($n$ = 5, 5, 10, 6, 6 for each group, respectively; the same $n$ applies to panels b-e). (b) Scatter plot of the long-time MSD exponent $\gamma_2$ ($p$ = $2.1 \times 10^{-10}$). (c) Self diffusivity $D_s$ for different $R$ ($p$ = $7.6 \times 10^{-7}$). (d) Steady-state cell shape index $q_\text{steady}$ ($p$ = $4.1 \times 10^{-7}$). (e) Root mean square traction $T$. Error bars denote standard deviations. Horizontal lines indicate means. All $p$-values were obtained from two-sided linear correlation tests against the null hypothesis of no correlation. The fluidity of the co-culture increases rapidly with the KO fraction ($R$), with diffusivity increasing by nearly two orders of magnitude while the changes in $q_\text{steady}$ are negligible. This observation is consistent with the DECMA-1 and EGTA perturbations, though interpretation of these results are complicated by the fact that the E-cadherin KO cells generate higher traction forces on the substrate (consistent with earlier studies~\cite{balasubramaniam2021investigating}). Nevertheless, the overall trends remain consistent across all perturbations: reduced cell-cell adhesion enhances tissue fluidity with minimal changes in cell shape.}
\label{Co-culture_Exp}
\end{figure}

\begin{figure}[!htbp]
\centering
\includegraphics[width=6.5in]{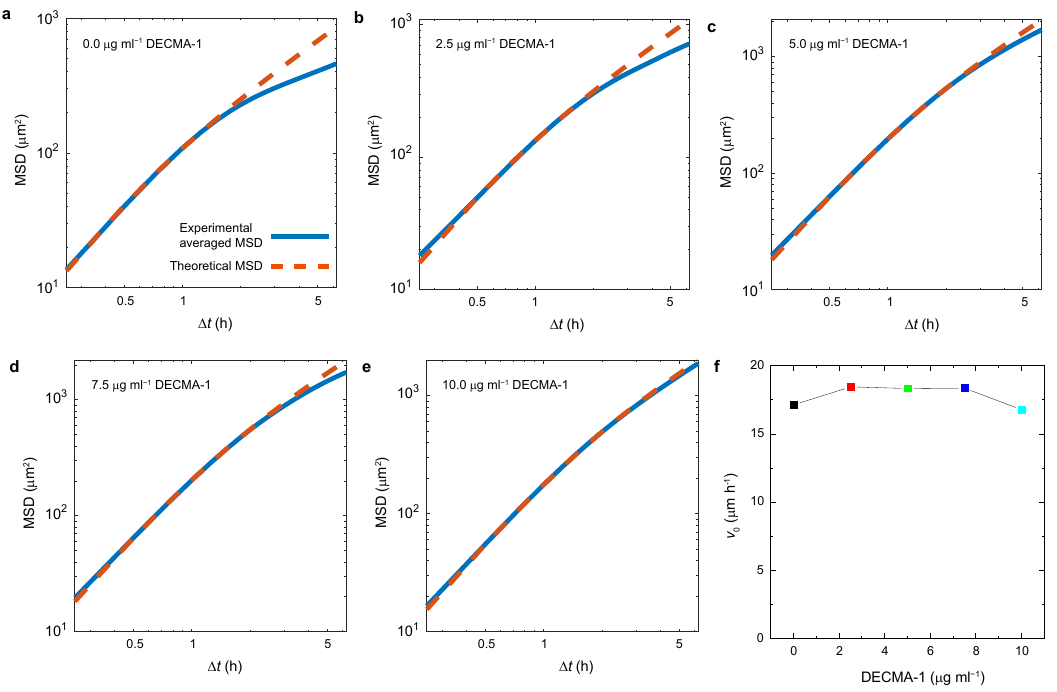}
\caption{MSD of cells at different adhesion levels (DECMA-1 concentrations) are fitted to the theoretical MSD of self-propelled particles undergoing rotational diffusion (main text Eqn.~5). Fits are shown separately in panels (a-e) for all DECMA-1 concentrations. (f) Extracted self-propulsion strength $v_0$ plotted against DECMA-1 concentration.}
\label{prw_msd}
\end{figure}

\clearpage 

\section*{Supplementary Tables}

\renewcommand{\tablename}{Supplementary Table}
\renewcommand{\thetable}{\arabic{table}}
\setcounter{table}{0}

\newcolumntype{C}{>{\centering\arraybackslash}X}

\begin{table}[!htbp]
\caption{Extended statistical analysis of results in Fig.~2 (DECMA-1 treatment) using one-way ANOVA followed by two-sided Dunnett's multiple-comparisons ($n$ = 4 per group). Adjusted $p$ values are reported relative to the sham control (0~\textmu g~ml$^{-1}$).}
\centering
\footnotesize
\begin{tabularx}{0.28\linewidth}{|c|c|c|}
\hline
DECMA-1 & $v_\mathrm{IR}$ & $T_\text{steady}$  \\
(\textmu g~ml$^{-1}$) & $p$-value & $p$-value \\
\hline
2.5 & 0.045 & 1.0 \\
5   & 0.66 & 0.96 \\
7.5 & 0.67 & 0.95 \\
10  & 0.96 & 0.73 \\
\hline
\end{tabularx}
\end{table}

\begin{table}[!htbp]
\caption{Extended statistical analysis of results in Fig.~3 (EGTA treatment) using one-way ANOVA followed by two-sided Dunnett's multiple-comparisons ($n$ = 4, 3, 5, 5, 6, 6 for each group, respectively). Adjusted $p$ values are with respect to sham control (0~mM).}
\centering
\footnotesize
\begin{tabularx}{0.33\linewidth}{|c|c|c|c|}
\hline
EGTA & $q_\text{steady}$ & $v_\mathrm{IR}$ & $\rho_\text{steady}$ \\
(mM) & $p$-value & $p$-value & $p$-value \\
\hline
0.5  & 1.0 & 0.020 & 1.0 \\
1    & 0.99 & 0.040 & 0.014 \\
1.5  & 1.0 & 0.95 & 0.92 \\
1.7  & 0.46 & 0.31 & 0.60 \\
1.8  & 0.36 & 0.37 & 0.76 \\
\hline
\end{tabularx}
\end{table}

\begin{table}[!htbp]
\caption{Model parameters that best match the experiments.
See ``Estimating model units and parameters'' in Methods for the boundary conditions, a quantitative estimate of boundary effects, and a detailed description of simulation parameters and constraints. The parameters in this table correspond to the stars in Fig.~4d. The dimensionless quantities can be computed using the unit of time as $t_0$ = $\mu/(K_A \bar{A})$, the unit of length as $l_0$ = $\sqrt{\bar{A}}$, and the unit of force as $K_A\bar{A}^{3/2}$. As described in the main text, the dimensionless viscous drag is given by $\tilde{\eta}$ = $\eta\sqrt{\bar{A}}/\mu$. }
\centering
\footnotesize
\begin{tabularx}{0.82\linewidth}{|c|c|c|c|c|c|c|}
        \hline
        DECMA-1               & cell-substrate                & area                        & average cell              & preferred   & junctional \\
        concentration         & viscous coefficient           & elasticity                  & area                      & shape index & drag coefficient \\
        (\textmu g~ml$^{-1}$) & $\mu$ (nN.h~\textmu m$^{-1}$) & $K_A$ (nN~\textmu m$^{-3}$) & $\bar{A}$ (\textmu m$^2$) & $p_0$       & $\eta$ (nN.h~\textmu m$^{-2}$) \\
        \hline
        0   & 0.247 & 0.11 & 349 & 3.88 & 0.126\\
        2.5 & 0.247 & 0.11 & 391 & 3.89 & 0.083\\
        5   & 0.247 & 0.11 & 434 & 3.91 & 0.048\\
        7.5 & 0.247 & 0.11 & 430 & 3.91 & 0.043\\
        10  & 0.247 & 0.11 & 433 & 3.90 & 0.027\\
        \hline
\end{tabularx}
\end{table}

\begin{table}[!th]
\caption{Extended statistical analysis of results in Supplementary Fig.~4 (DECMA-1 treatment) using one-way ANOVA followed by two-sided Dunnett's multiple-comparisons ($n$ = 4 per group). Adjusted $p$ values are with respect to sham control (0~\textmu g~ml$^{-1}$).}
\centering
\footnotesize
\begin{tabularx}{0.2\linewidth}{|c|c|}
\hline
DECMA-1 & $q_\text{steady}$ \\
(\textmu g~ml$^{-1}$) & $p$-value \\
\hline
2.5 & 0.059 \\
5   & 0.017 \\
7.5 & 0.019 \\
10  & 0.24 \\
\hline
\end{tabularx}
\end{table}

\begin{table}[!th]
\caption{Extended statistical analysis of results from the DECMA-1 experiment at higher cell density shown in Supplementary Fig.~5 using one-way ANOVA followed by two-sided Dunnett's multiple-comparisons ($n$ =  6, 6, 9, 5 for each group, respectively). Adjusted $p$ values are in comparison to 0~\textmu g~ml$^{-1}$ DECMA-1 (control + sham control).}
\centering
\footnotesize
\begin{tabularx}{0.28\linewidth}{|c|c|c|}
\hline
DECMA-1 & $D_s$ & $q_\text{steady}$ \\
(\textmu g~ml$^{-1}$) & $p$-value & $p$-value \\
\hline
1.3 & 0.020 & 0.0073 \\
2.5 & 0.051 & 0.023 \\
5 & 0.32 & 0.61 \\
\hline
\end{tabularx}
\end{table}

\end{document}